\documentclass[aps,prd,twocolumn,preprintnumbers,showpacs,nofootinbib,amssymb]{revtex4}
\usepackage{graphicx}
\usepackage{amsmath}
\usepackage{amssymb}
\usepackage{amsfonts}
\usepackage{bm}
\usepackage{color}

\def\be{\begin{equation}}
\def\ee{\end{equation}}
\def\bea{\begin{eqnarray}}
\def\eea{\end{eqnarray}}

\begin{document}

\title{Derivation of a generalized Schr\"odinger equation for dark matter
halos\\
 from the theory of scale relativity}
\author{Pierre-Henri Chavanis}
\affiliation{Laboratoire de Physique Th\'eorique,
Universit\'e Paul Sabatier, 118 route de Narbonne  31062 Toulouse, France}

\begin{abstract} 

Using Nottale's theory of scale relativity, we derive a generalized
Schr\"odinger equation applying to dark matter halos. This equation
involves a logarithmic nonlinearity associated with an effective temperature and a source of dissipation.
Fundamentally, this wave equation arises from the nondifferentiability of the 
trajectories of the dark matter particles whose origin may be due to ordinary
quantum mechanics, classical ergodic (or
almost ergodic) chaos, or to the fractal nature of
spacetime at the cosmic scale.
The generalized Schr\"odinger equation involves a coefficient ${\cal D}$,
possibly different from $\hbar/2m$ (where $\hbar$ is the Planck constant and
$m$ the mass of the particles), whose value for dark matter
halos is ${\cal D}=1.02\times 10^{23}\, {\rm m^2/s}$. This model is
similar to the Bose-Einstein
condensate dark matter model except 
that it does not require the dark matter particle to be ultralight. It can
accomodate any type of particles provided that they have
nondifferentiable trajectories. We 
suggest that the cold dark matter crisis may be solved by the fractal
(nondifferentiable) structure of spacetime at the cosmic scale, or by the
chaotic motion of the particles on a very long timescale, instead of
ordinary quantum mechanics. The equilibrium states of the generalized
Schr\"odinger equation correspond to 
configurations with a core-halo structure. The quantumlike potential generates a
solitonic core that solves the
cusp problem of the classical cold dark matter model. The logarithmic
nonlinearity
accounts for the presence of an isothermal halo that leads to flat rotation
curves (it also accounts for the isothermal core of large dark matter
halos). The damping term ensures that the system relaxes towards an
equilibrium
state. This property is guaranteed by an $H$-theorem satisfied by a
Boltzmann-like free
energy functional.
In our
approach, the temperature and the friction arise
from a single formalism. They correspond to the real and imaginary parts of the
complex friction coefficient present in the scale covariant equation 
of dynamics that is at the basis of Nottale's theory of scale relativity.  They
may be the
manifestation of a cosmic aether  or the consequence
of a process of
violent
relaxation and gravitational cooling on a coarse-grained scale. 
 
\end{abstract}


\maketitle

\section{Introduction}
\label{sec_introduction}

The nature of dark matter (DM) is still unknown and remains one of the most
important open problems
of
modern cosmology. The existence of DM has been predicted by Zwicky in 1933 to
account for the missing mass of the galaxies in the Coma cluster inferred from the virial theorem
\cite{zwicky}. The robust indication of DM came later from the measurement of 
the rotation curves of spiral galaxies \cite{flat1,flat2,flat3,flat4} that
revealed
that they were flat instead of declining with the distance like in the case of
planetary systems that are dominated by a central mass (Kepler's law). The
existence of DM has
been confirmed by independent observations of gravitational lensing
\cite{massey}, hot gas in clusters \cite{hotgas}, and the anisotropies of the
cosmic microwave background (CMB) \cite{cmb}. Very recently, astronomers
reported that Dragonfly 44, an ultra diffuse galaxy with the mass of the Milky
Way but with no discernable stars may be made almost entirely of DM
\cite{dragonfly}.

Observation of the large-scale
structure (LSS) of the Universe and the CMB are
consistent with the cold dark matter (CDM) model in which DM is
modeled as
a pressureless gas described
by the Euler-Poisson equations or as a collisionless system described
by the Vlasov-Poisson equations. The most
studied candidate particles for DM are WIMPs (weakly interacting massive
particles) with a mass in the GeV-TeV range. These particles are the lightest
supersymmetric partners predicted by models of supersymmetry (SUSY)
\cite{susy}. The CDM model works remarkably well at
large (cosmological) scales and is consistent with ever improving measurements
of the CMB from WMAP and Planck missions
\cite{planck2013,planck2015}. It is able to account for the
formation of structures, with the small objects forming first and merging over
time to form larger objects (hierarchical clustering). This leads to a ``cosmic
web'' made of virialized halos connected by filaments delimiting empty voids, in
very good agreement with observations. 
However, the CDM model experiences serious difficulties at small (galactic)
scales. In particular, being pressureless, numerical simulations of CDM lead to
cuspy density profiles \cite{nfw}, with a density diverging as
$r^{-1}$ for $r\rightarrow 0$,  while observations favor cored density
profiles with a finite density at the center \cite{observations}. This is the
``cusp-core'' problem \cite{cusp}. Other related problems are known as the
``missing
satellites'' problem \cite{satellites1,satellites2,satellites3,satellites4} and
the ``too big to fail''
\cite{tbtf} problem. The
expression 
``small-scale crisis of CDM'' has been coined.\footnote{Some
researchers remain  unconvinced that there is a real problem at the center of
the
galaxies. For example, the cusp-core problem could be an effect of asphericity
and misalignment of the halos. We refer to \cite{genina} for a detailed
discussion of this issue and additional references.}

One possibility to solve the CDM crisis is to invoke the feedback
of baryons
that can transform cusps into cores \cite{romano1,romano2,romano3}. Another
possibility is to
consider
warm dark matter (WDM) where the dispersion of the particles is responsible for
a pressure force that can halt gravitational collapse and prevent the formation
of cusps \cite{wdm}. 
Finally, an interesting suggestion is to invoke quantum mechanics. Indeed, DM
could be made of elementary particles that have not been detected yet and whose
quantum nature may solve the small-scale problems of the CDM model.

For example, the DM
particle could be a fermion, like a sterile neutrino with a mass in the keV
range,  satisfying the Fermi-Dirac statistics (a
sterile neutrino is also the most plausible
candidate for WDM). In that case,
gravitational collapse leading to cuspy halos is
prevented by the quantum pressure arising from the Pauli exclusion principle, or
by the thermal pressure. The
resulting configurations have a
core-halo structure with a core made of a
``fermion ball'' at $T=0$  and an isothermal halo with a density profile
decaying
as $\rho\sim r^{-2}$ at large distances.\footnote{This  $\rho\sim
r^{-2}$ profile is similar to
the
numerical Navarro-Frenk-White (NFW) \cite{nfw} profile and to the observational
Burkert \cite{observations} profile both decaying as $\rho\sim
r^{-3}$ at large distances. The difference of slope between the isothermal
profile  ($\rho\sim r^{-2}$) and the NFW and Burkert profiles 
($\rho\sim r^{-3}$)  may be due to nonindeal effects: incomplete
relaxation, tidal effects, stochastic forcing...} An isothermal, or
almost isothermal halo, leads to flat rotation cuves. 
This
core-halo structure with a degenerate core and an isothermal halo may also be
justified by
the process of collisionless violent relaxation (for classical or quantum
particles) leading to the Lynden-Bell statistics \cite{lb} that is similar to
the
Fermi-Dirac statistics. This violent collisionless relaxation,
establishing an
out-of-equilibrium Fermi-Dirac-like distribution on a very short timescale, may
be more relevant than a collisional relaxation establishing a thermal
Fermi-Dirac
distribution on a much longer timescale, possibly larger than the age of the
Universe \cite{csmnras,clm1,clm2}.
This
fermionic scenario, where the Fermi-Dirac distribution arises either from
quantum mechanics or from violent collisionless relaxation, has been
studied by several
authors \cite{markov,cmc1,cmc2,gao,stella,zcls,cls,cl,ir,gmr,merafina,imrs,vtt,
bvn,bmv,
csmnras,bvr,pt,dark,bmtv,btv,rieutord,ijmpb,dvs1,dvs2,vss,urbano,rar,clm1,clm2,
vs1,vs2,rsu,krut}.

Alternatively, the DM particle could be a boson like an axion. At low
temperatures, bosons are expected to  
form Bose-Einstein condensates (BECs). In that case, DM is described by a
scalar field (SF) that can be identified with the wave function $\psi$ of the
condensate. The evolution of this wave function is governed by the
Schr\"odinger-Poisson equations.
In the BEC scenario, gravitational collapse is prevented by the
repulsive
quantum potential accounting for the Heisenberg uncertainty principle. 
The resulting structure has a core-halo structure. Quantum mechanics is important
in the core which is similar to a soliton (a steady state of the
Schr\"odinger-Poisson equations). On the other hand, the halo
(quantum interferences)
behaves essentially as CDM and has a density profile close to the isothermal
profile ($\rho\sim r^{-2}$) or close to the  
NFW and Burkert  profiles ($\rho\sim r^{-3}$). This core-halo structure may
result from a process of gravitational
cooling \cite{seidel94}  that is 
similar to the process of violent relaxation \cite{lb}. Gravitational cooling
may be at work during hierarchical clustering where DM halos merge to form
bigger halos. This bosonic model has received different names such as wave DM,
fuzzy DM (FDM), BECDM, $\psi$DM, or
SFDM \cite{baldeschi,khlopov,membrado,sin,jisin,leekoh,schunckpreprint,
matosguzman,sahni,
guzmanmatos,hu,peebles,goodman,mu,arbey1,silverman1,matosall,silverman,
lesgourgues,arbey,fm1,bohmer,fm2,bmn,fm3,sikivie,mvm,lee09,ch1,lee,prd1,prd2,
prd3,briscese,
harkocosmo,harko,abrilMNRAS,aacosmo,velten,pires,park,rmbec,rindler,lora2,
abrilJCAP,mhh,lensing,glgr1,ch2,ch3,shapiro,bettoni,lora,mlbec,madarassy,
abrilph,playa,stiff,guth,souza,freitas,alexandre,schroven,pop,eby,cembranos,
braaten,davidson,schwabe,fan,calabrese,marsh,bectcoll,cotner,chavmatos,hui,
tkachevprl,abrilphas,shapironew,total,moczetal,phi6,
abriljeans,zhang,moczchavanis,desjacques} (see the
introduction of
\cite{prd1} for a short historic of this model). The
relevance of this model has been demonstrated 
by the simulations of Schive {\it et al.} \cite{ch2,ch3}. They showed
that the BECDM model behaves as 
CDM at large (cosmological) scales but that differences appear at small
(galactic) scales where the wave nature of the 
particles manifests itself. This may solve the CDM crisis. For
quantum
mechanics to be important at the scale of DM halos, the mass of the boson
must be extremely small, of the order of $10^{-22}\, {\rm eV/c^2}$ (this is the
condition
required for the de Broglie wavelength of the boson to be of the order of the
size of  DM halos).
The standard QCD axion with a mass $m=10^{-4}\, {\rm
eV/c^2}$ essentially behaves as CDM and does not solve the CDM crisis. However,
ultralight axions with a mass up to $m\sim H_0\hbar/c^2\sim 10^{-33}\, {\rm
eV/c^2}$ (where $H_0$ is the Hubble constant) arise in string theory and are not
excluded by particle physics. This point has been recently
emphasized by Hui {\it et al.} \cite{hui} who stressed the viability
of the FDM model. Nevertheless, the existence of ultralight bosons remains an
hypothesis
that is not confirmed yet. It may therefore be interesting to develop
alternative
models of DM that do not require such small particle masses while exhibiting
features similar to the FDM model.

In this paper, we approach the problem of DM from the viewpoint
of Nottale's theory of scale relativity \cite{nottale} relying on a fractal
spacetime.\footnote{The theory of Nottale is related to, but distinct from,
Nelson's stochastic interpretation of quantum mechanics \cite{nelson}.} Nottale
has shown, quite generally, that when a particle has a nondifferentiable
trajectory, its
evolution is described by a Schr\"odinger-like equation. This equation
involves a coefficient ${\cal D}$ with the dimension of a diffusion coefficient, sometimes
called the fractal fluctuation parameter,  whose value depends upon the system
under consideration (i.e the origin of the nondifferentiability).

Nottale first argued that the fractal structure of spacetime
manifests itself at the ``microscale'' which is the realm of ordinary quantum
mechanics.
He used his formalism to derive the Schr\"odinger equation for a quantum
particle from Newton's fundamental equation of dynamics by invoking a principle
of
scale covariance. In order to reproduce
the results of quantum mechanics,
the coefficient  ${\cal D}$ which appears in his Schr\"odinger-like equation must be equal to ${\cal D}=\hbar/2m$, where
$\hbar$ is the Planck constant and $m$ is the mass of the particle.

Then, Nottale proposed that the fractal structure of spacetime also manifests
itself at the ``macroscale'' which is the realm of astrophysics and cosmology.
He writes: ``In this new approach, space becomes not only curved but also fractal
beyond some characteristic scale relative to the system under consideration. The
induced effects on motion (in standard space) of the internal fractal
structures of geodesics (in scale space), are to
transform classical mechanics
into a quantum-like mechanics, i.e. Newton's fundamental equation of dynamics
into a Schr\"odinger-like equation.'' At  the  macroscale,
the nondifferentiability of the trajectories of the particles could arise either
from the
chaoticity of the motion of the classical particles on a very long timescale
that is
larger than their predictability horizon (ergodic or almost ergodic chaos), or
from the intrinsic fractal structure of spacetime itself above a certain
astrophysical scale. In other words, spacetime can become fractal beyond some
temporal and/or space transition. This leads to a new quantum
mechanics operating now at the cosmic scale.

Nottale tried to find evidence of his theory in some astrophysical
observations. He first  applied his theory to the solar system
based on the fact that the planets have a chaotic dynamics on a long timescale. 
In his approach, the
solar system may be viewed as a gigantic atom described by a
Schr\"odinger-like equation with an attractive  $1/r$
potential.\footnote{Historically, the idea to describe the
solar system by a Schr\"odinger 
equation with a Newtonian gravitational potential dates back to Jehle
\cite{jehle} in 1938.} This leads to a
quantization of the solar system similar to the quantization of the hydrogen
atom. The
difference with ordinary quantum mechanics is that the
coefficient ${\cal D}$ that appears in the Schr\"odinger-like equation of Nottale  has a value different from
$\hbar/2m$. From this Schr\"odinger-like
equation, he could obtain a  quantization of the semi-major axes and
eccentricities of planetary orbits. The predicted law of distance is not a
Titius-Bode-like power law ($a_n=a+b\, c^n$) but a more constrained and
statistically significant quadratic law of the form $a_n=a_0 n^2$
(in these expressions $a_n$ is the semi-major
axis and $n>0$ is an integer quantumlike number).
Interestingly, this law gives a
much better fit to the planetary distribution than the empirical
Titius-Bode law.

Nottale applied his formalism to other
astrophysical objects such as extra-solar planetary systems, star-forming regions, binary stars,
high-velocity clouds, planetary nebulae, galactic centers, galaxies, our Local
Group of galaxies, clusters of galaxies, and the large scale structures of the
Universe.\footnote{He also applied his approach to the morphogenesis of
planetary nebulae (and even flowers!), obtaining
particular
solutions of the Schr\"odinger
equation that spectacularly resemble to these objects. It is
however
difficult to say whether this impressive resemblance is just a coincidence or if
this agreement is deeper than apparent at first sight.} In all these examples,
gravity acts as an
external potential that is not affected by the structures that it contributes to
form.

At last, Nottale proposed to apply his formalism to DM halos with, again, the
argument that the DM particles have  nondifferentiable trajectories due to chaos
or due to the fractal structure of spacetime at astrophysical scales. In the
case
of DM halos, the gravitational potential is produced by the system
itself in a self-consistent manner. As a result, Nottale obtained a
Schr\"odinger-Poisson-like
equation  similar to the Schr\"odinger-Poisson equation that arises in the
BEC/SF model of
DM.\footnote{Nottale did not point out this analogy probably because he did
not know the literature on this subject.} Again, the main difference is that
the coefficient ${\cal D}$ in Nottale's theory may be different from
$\hbar/2m$.

This suggestion is very interesting because it could give a novel
interpretation to the Schr\"odinger-Poisson equation applying to DM halos, 
different from its interpretation
stemming from ordinary quantum mechanics. However, some arguments given by
Nottale are incorrect. For example, Nottale argues
that there is no need of DM to explain the flat
rotation curves of spiral galaxies. He writes: ``The quantumlike potential $V_Q$
is at the
origin of the various dynamical and lensing effects usually attributed to unseen
additional masses''.

Based on the results obtained by solving the Schr\"odinger-Poisson equation in
the context of the
BEC/SF model of DM \cite{prd1}, we know that the
effect of the quantum potential is different from what is suggested by Nottale.
Still, it plays a fundamental role in the physics of DM halos 
since it can solve the cusp problem. This leads to the first potentially important result
of this paper. We propose that the cusp problem may be solved by the quantumlike
potential arising in the theory of Nottale from the nondifferentiability of the trajectories of
the DM particles. This  nondifferentiability may be due to (i) the
quantum nature of the particles if they are ultralight (as in the usual
interpretation of the FDM model), (ii) their chaotic (fractal) dynamics that
manifests itself on a very
long timescale, or (iii)
the intrinsic fractal structure of spacetime above a certain
scale.\footnote{If the nondifferentiability of spacetime and
the Schr\"odinger-like equation are due to an effect of chaos in classical
mechanics, we may wonder why cores are not formed in $N$-body
simulations. A
possibility is that the simulations (being necessarily based on approximations)
are not fully reliable over long times to account for subtle effects of chaos.
Inversely, assuming that the  $N$-body simulations are fully reliable  would
imply that possibility (ii) has to be rejected. This would leave us, in the
framework of our approach, with possibility (i) involving ordinary quantum
mechanics or
with possibility (iii) involving interesting new physics.} If
suggestions (ii) and/or (iii) are correct, that
would mean that the cored density profile of DM halos is a manifestation of
the fractal  nature of spacetime at astrophysical
scales. That would lead to a revolution of the concepts of space and time 
since one would have to take into account the fractal
nature of spacetime in the theories of physics 
when dealing with the large-scale structures of the Universe or considering very
long timescales of evolution.

That would also have important implications for the 
nature of the DM particle. Indeed, if the cored density profile of DM halos is
due to the quantization of the Universe at the macroscale instead of being due
to ordinary quantum mechanics, the observation of cored
density profiles should be independent of the mass of the DM particle in
agreement with the equivalence principle.
Indeed, 
observations of DM halos only determine the coefficient ${\cal D}$ in the
Schr\"odinger-like equation (see below), and this value may be different
from $\hbar/2m$. That would enlarge the possibility of particles composing DM.
Ultralight axions with a mass of the order of $10^{-22}\, {\rm eV/c^2}$ are not
required anymore. Cores, instead of cusps, could be obtained with more massive
bosons such as the QCD axion or even with classical particles such as WIMPs
provided that the quantization
of the DM halos (quantum potential) comes from the intrinsic fractal nature of
spacetime or from the chaoticity of the trajectories of the particles.

In this paper, we complement Nottale's theory of scale relativity by considering
a more
general situation where the particles are submitted to dissipative effects in addition to Newton's law.
The origin of this dissipation may be due to (i) the interaction of the system
with an external environment (a real one or an hypothetical aether), (ii)  the
complex evolution of the system itself
(gravitational cooling or violent relaxation) that leads to an effective
dissipation on the coarse-grained scale, or
(iii) the intrinsic nature of the
fractal spacetime. We derive from the theory of scale relativity a generalized
Schr\"odinger equation that involves a logarithmic nonlinearity associated with
an effective temperature and a
source of dissipation. In our approach, the temperature and the friction
arise from a single formalism. They correspond
to the real and imaginary parts of the complex friction coefficient present in
the scale covariant equation of dynamics that is at the basis of Nottale's
theory of scale relativity. When applied to DM halos, our generalized
Schr\"odinger equation has
interesting properties. Its equilibrium states
have a core-halo structure. The quantumlike potential accounts for the
solitonic
core of DM halos solving the cusp problem. The logarithmic nonlinearity
accounts for their isothermal halo leading to flat rotation
curves (it also accounts for the isothermal core of large DM
halos). The damping term ensures that the system relaxes towards this
equilibrium state. This is guaranteed by an $H$-theorem satisfied by a
Boltzmann-like
free energy functional.
We use observations of DM halos to
determine the coefficient ${\cal D}$ arising in the Schr\"odinger-like
equation and find ${\cal D}=1.02\times 10^{23}\, {\rm
m^2/s}$.

\section{Derivation of a generalized Schr\"odinger equation}
\label{sec_derivation}

\subsection{Basic tools of scale relativity}
\label{sec_basic}

When a particle has a nondifferentiable trajectory  ${\bf r}(t,dt)$, the
derivative $d{\bf r}/dt$ is not defined and one has to introduce
two velocities ${\bf u}_+({\bf r}(t),t)$ and  ${\bf u}_-({\bf r}(t),t)$ defined
from $t-dt$ to $t$ for ${\bf u}_-$ and from $t$ to $t+dt$ for ${\bf u}_+$
\cite{nottale}.
The two-valuedness of the velocity vector is due to the irreversibility in the
reflection $dt\leftrightarrow -dt$ (time symmetry breaking). The
elementary displacement $d{\bf r}_{\pm}$ for both processes  is the sum of a
differential part $d{\bf r}_{\pm}={\bf u}_{\pm}\, dt$ and a non-differentiable
part which is a scale-dependent fractal fluctuation $d{\bf b}_{\pm}$. This
fractal fluctuation is described by a stochastic variable which, by definition,
is of zero mean $\langle d{\bf b}_{\pm}\rangle ={\bf 0}$. We shall assume that
spacetime has a fractal dimension $D_{\rm F}=2$ as in ordinary 
quantum mechanics \cite{feynman}. More general models  could be
constructed  if the fractal dimension is different from $D_{\rm F}=2$
\cite{nottale} but, in
this paper, we restrict ourselves to the simplest case.
Therefore, we write
\begin{equation}
\label{b1}
d{\bf r}_{\pm}={\bf u}_{\pm}\, dt+d{\bf b}_{\pm},
\end{equation}
with
\begin{equation}
\label{b2}
\langle d{\bf b}_{\pm}\rangle ={\bf 0},\qquad \langle db_{\pm i} db_{\pm
j}\rangle =\pm 2{\cal D}\delta_{ij}dt,
\end{equation}
where the indices $i,j$ denote the
coordinates of space and ${\cal D}$ is a a sort of ``diffusion coefficient''
measuring the
covariance of the noise.\footnote{In Eq. (\ref{b2}), we consider that $dt>0$ for
the $(+)$ process and
$dt<0$ for the $(-)$ process so that $\pm 2{\cal D}\delta_{ij}dt$ is always
positive \cite{nottale}.}  In other
words, ${\cal D}$ characterizes the amplitude of the fractal fluctuations.
Following Nottale, we introduce two classical derivative operators
$d_+/dt$ and $d_-/dt$
which yield the twin classical velocities when they are applied to the position
vector ${\bf r}$, namely
\begin{equation}
\label{b3}
\frac{d_+{\bf r}}{dt}={\bf u}_+,\qquad \frac{d_-{\bf r}}{dt}={\bf u}_-.
\end{equation}
It proves convenient in the formalism to replace the twin velocities $({\bf
u}_+,{\bf u}_-)$ by the couple $({\bf u}, {\bf u}_Q)$ where
\begin{equation}
\label{b4}
{\bf u}=\frac{{\bf u}_++{\bf u}_-}{2},\qquad {\bf u}_Q=\frac{{\bf u}_+-{\bf
u}_-}{2}.
\end{equation}
With these two velocities, we can form a complex velocity
\begin{equation}
\label{b5}
{\bf U}={\bf u}-i{\bf u}_Q.
\end{equation}
The real part ${\bf u}$ can be identified  with the classical velocity and the
imaginary part ${\bf u}_Q$ is a manifestation of the nondifferentiability of
spacetime. It will be
called the quantum velocity. For a differentiable
trajectory ${\bf u}_+={\bf u}_{-}={\bf u}$ and ${\bf u}_Q={\bf 0}$. As we
shall see, the complex velocity ${\bf U}$ leads to the
Schr\"odinger
equation. Therefore, the origin
of complex numbers in the  Schr\"odinger equation (the wave function $\psi$ and
the complex number $i$) can be intrinsically attributed to the two-valuedness character
of the velocity \cite{nottale}.

Following Nottale, we
 define a complex
derivative operator from the classical (differential) parts as
\begin{equation}
\label{b6}
\frac{D}{Dt}=\frac{d_++d_-}{2dt}-i \frac{d_+-d_-}{2dt}
\end{equation}
in terms of which
\begin{equation}
\label{b7}
\frac{D{\bf r}}{Dt}={\bf U}.
\end{equation}
The total derivative with respect to time of a function $f({\bf r}(t),t)$ of
fractal dimension $D_{\rm F}=2$ writes
\begin{equation}
\label{b8}
\frac{df}{dt}=\frac{\partial f}{\partial t}+\nabla f\cdot \frac{d{\bf
r}}{dt}+\frac{1}{2}\sum_{i,j}\frac{\partial^2 f}{\partial x_i\partial
x_j}\frac{dx_i dx_j}{dt}.
\end{equation}
Using Eq. (\ref{b2}), we
find that the classical (differentiable) part of this expression is
\begin{equation}
\label{b9}
\frac{d_{\pm}f}{dt}=\frac{\partial f}{\partial t}+{\bf u}_{\pm}\cdot \nabla f\pm
{\cal D}\Delta f.
\end{equation}
Substituting Eq. (\ref{b9}) into Eq. (\ref{b6}), we obtain the expression of the
complex time derivative operator \cite{nottale}:
\begin{equation}
\label{b10}
\frac{D}{Dt}=\frac{\partial}{\partial t}+{\bf U}\cdot \nabla-i {\cal D}\Delta.
\end{equation}

\subsection{Application to self-gravitating systems}

We now apply this formalism to a system of $N$ nonrelativistic particles
of mass $m$ in gravitational interaction. If
their trajectories are differentiable, each particle has an equation of motion
given by Newton's law 
\begin{equation}
\label{qb1z}
\frac{d{\bf u}}{dt}=-\nabla\Phi,
\end{equation}
where ${\bf F}=-\nabla\Phi$ is the gravitational force by unit of mass exerted on the
particle. We note that the mass $m$ of the particles does not appear in 
this equation in virtue of the equivalence principle.
 If we make a meanfield
approximation, valid for $N\rightarrow +\infty$ with $m\sim 1/N$, the
gravitational potential $\Phi({\bf r},t)$ can be identified with the
self-consistent potential produced by the system as  whole $\Phi({\bf
r},t)=-G\int \rho({\bf
r}',t)/|{\bf r}-{\bf r}'|\,
d{\bf r}'$ through the Poisson
equation
\begin{equation}
\label{poisson}
\Delta\Phi=4\pi G\rho,
\end{equation}
where $\rho({\bf r},t)$ denotes the density of particles. In that case, the
evolution of the
distribution function $f({\bf r},{\bf v},t)$ in phase space is governed by the
Vlasov equation
which describes the collisionless evolution of the system \cite{bt}. If we
take finite-$N$ effects into account (gravitational encounters), we obtain at
the order $1/N$
the inhomogeneous Lenard-Balescu equation \cite{heyvaerts,angleaction} which
describes the collisional
evolution of the system on a secular timescale. This equation
can be derived rigorously from the $N$-body equations of motion in a systematic
expansion in powers of $1/N$.

In this paper, we make a mean field approximation ($N\rightarrow +\infty$) but
we consider the
possibility that the trajectories of the particles are
nondifferentiable for one of the reasons given in the Introduction (ordinary
quantum mechanics, chaos, fractal nature of spacetime...). 
 We use the fundamental postulate of
Nottale's theory of scale relativity according to which the equations of quantum mechanics (nondifferentiable trajectories) can be
obtained from the equations of classical mechanics (differentiable trajectories)
by replacing the standard velocity ${\bf u}$ by the complex velocity ${\bf U}$
and the standard time derivative $d/dt$ by the complex time derivative $D/Dt$.
In other words, $D/Dt$ plays the role of a ``covariant derivative operator'' in
terms of which the fundamental equations of physics keep the same form in the
classical (differentiable) and quantum (nondifferentiable) regimes.\footnote{In
the present context, the term ``quantum'' should be taken in a very general
sense, valid either at the microscale (ordinary quantum mechanics)  or at the
macroscale (new quantum mechanics).} This is similar to the principle of
covariance in
Einstein's theory of relativity according to which the form of the equations of
physics should be conserved under all transformations of coordinates.

\subsection{Complex friction force}

In order to be general, we assume that the particles are
submitted to a friction force in addition to self-gravity. Introducing a source
of dissipation in the fundamental equation of dynamics is the next
level of complexity after the pure Newton law (\ref{qb1z}). The naive idea is
to introduce a
linear complex friction force of the form $-\gamma {\bf U}$, where $\gamma$ is a
complex friction coefficient. However, it proves
necessary to consider only the real part of the friction force in order to
obtain a generalized Schr\"odinger equation that conserves the normalization
condition locally (see Appendix F of \cite{chavnot}). Therefore, we write the
scale
covariant equation of dynamics
under the form
\begin{equation}
\label{qb1}
\frac{D{\bf U}}{Dt}=-\nabla\Phi-{\rm Re} (\gamma {\bf U}).
\end{equation}
Using
the expression (\ref{b10}) of the complex time
derivative
operator,
the foregoing equation can be rewritten as
\begin{equation}
\label{qb2}
\frac{\partial {\bf U}}{\partial t}+({\bf U}\cdot \nabla){\bf U}=i {\cal
D}\Delta{\bf U}-\nabla\Phi-{\rm Re} (\gamma {\bf U}).
\end{equation}
This equation is similar to the damped viscous Burgers equation of fluid mechanics,
except that in the present case the velocity field ${\bf U}({\bf r},t)$ is
complex and the
viscosity $\nu=i{\cal D}$ is imaginary.  

We now assume that the flow is
potential so that the complex velocity can be written 
as the gradient of a function, ${\bf U}=\nabla\Sigma$, where 
 $\Sigma$ is a complex potential or a complex
action.\footnote{See \cite{nottale} for a justification of this
assumption from Lagrangian mechanics.} As a consequence, the
flow is irrotational: $\nabla\times {\bf U}={\bf 0}$. Using the well-known
identities
of fluid mechanics $({\bf U}\cdot \nabla){\bf U}=\nabla ({{\bf U}^2}/{2})-{\bf
U}\times (\nabla\times {\bf U})$ and $\Delta{\bf U}=\nabla(\nabla\cdot {\bf
U})-\nabla\times(\nabla\times{\bf U})$  which reduce to $({\bf U}\cdot
\nabla){\bf U}=\nabla ({{\bf U}^2}/{2})$ and $\Delta{\bf U}=\nabla (\nabla\cdot
{\bf U})$ for an irrotational flow, and using the identity $\nabla\cdot {\bf
U}=\Delta\Sigma$, we find that Eq. (\ref{qb2})
is equivalent to
\begin{equation}
\label{qb3}
\frac{\partial \Sigma}{\partial t}+\frac{(\nabla \Sigma)^2}{2}-i{\cal
D}\Delta\Sigma+\Phi+V(t)+{\rm Re} (\gamma \Sigma)=0,
\end{equation}
where $V(t)$ is a ``constant'' of integration depending on time.
Equation (\ref{qb3}) can be viewed as a quantum Hamilton-Jacobi equation 
for a complex action, or as a Bernoulli equation  for a complex potential.

We
define the wave function $\psi({\bf r},t)$ through the complex Cole-Hopf
transformation
\begin{equation}
\label{qb4}
\Sigma=-2i{\cal D}\ln\psi.
\end{equation}
Written under the form 
\begin{equation}
\psi=e^{i\Sigma/2{\cal D}},
\end{equation}
this relation is  similar 
to the WKB transformation in quantum mechanics. Substituting Eq. (\ref{qb4})
into Eq. (\ref{qb3}), and using the identity
\begin{equation}
\label{qb5}
\Delta(\ln\psi)=\frac{\Delta\psi}{\psi}-\frac{1}{\psi^2}(\nabla\psi)^2,
\end{equation}
we obtain the generalized
Schr\"odinger equation
\begin{equation}
\label{qb6}
i{\cal D}\frac{\partial\psi}{\partial
t}=-{\cal D}^2\Delta\psi+\frac{1}{2}\Phi\psi+\frac{1}{2}V(t)\psi+{\cal D} \,
{\rm
Im}(\gamma\ln\psi)\psi.
\end{equation}
Therefore, by performing the complex  Cole-Hopf transformation (\ref{qb4}), we
find that the complex (damped) viscous Burgers equation (\ref{qb2})  is
formally equivalent to the (generalized) Schr\"odinger equation  (\ref{qb6}) in
the same sense that, by performing the usual Cole-Hopf transformation, the
viscous Burgers equation is equivalent to the diffusion
equation in ordinary hydrodynamics. As a result, quantum mechanics may be
interpreted as a generalized hydrodynamics involving a complex velocity field
and an imaginary viscosity. This interpretation takes a clear meaning 
in the context of Nottale's theory of scale relativity.

As will be demonstrated below, the density $\rho$ is proportional to $|\psi|^2$.
For commodity, we choose to
normalize the wave function such that $\int |\psi|^2\, d{\bf r}=M$, where $M$ is
the total mass of the system. This implies that
\begin{equation}
\label{jeu}
\rho=|\psi|^2.
\end{equation}
As a result, the generalized Schr\"odinger equation  (\ref{qb6}) must be coupled
self-consistently to the Poisson equation 
\begin{equation}
\label{dw2}
\Delta\Phi=4\pi G |\psi|^2.
\end{equation}
Dividing Eq. (\ref{qb6}) by $\psi$, taking the Laplacian, and
using the Poisson equation (\ref{dw2}), we can eliminate the gravitational
potential and obtain the single differential equation 
\begin{equation}
\label{weqp}
i{\cal D}\frac{\partial\Delta\ln\psi}{\partial
t}=-{\cal
D}^2\Delta\left
(\frac{\Delta\psi}{\psi}\right )+2\pi G|\psi|^2+{\cal
D} \,
{\rm
Im}(\gamma\Delta\ln\psi).
\end{equation}

\subsection{Recovery of ordinary quantum mechanics}
\label{sec_rec}

Before going further, let us make the connection with ordinary quantum 
mechanics. In the absence of
dissipation  ($\gamma=V=0$), the wave equation (\ref{qb6}) reduces to the
Schr\"odinger-like equation
\begin{equation}
\label{qm1}
i{\cal D}\frac{\partial\psi}{\partial
t}=-{\cal D}^2\Delta\psi+\frac{1}{2}\Phi\psi.
\end{equation}
This equation coincides with the ordinary  Schr\"odinger equation of quantum mechanics
\begin{equation}
\label{qm2}
i\hbar\frac{\partial\psi}{\partial t}=-\frac{\hbar^2}{2m}\Delta\psi+m\Phi\psi
\end{equation}
provided that we make the identification
\begin{equation}
\label{qm3}
{\cal D}=\frac{\hbar}{2m},
\end{equation}
where $\hbar$ is the Planck constant and $m$ the mass of the particles.
Therefore, in the context of ordinary quantum mechanics, the coefficient ${\cal
D}$ is inversely proportional to the mass of the particles.

In the gravitational case considered here, the
Schr\"odinger-like
equation (\ref{qm1}) satisfies the equivalence
principle
since it does not
depend on the inertial mass $m$ of the particles. This is
consistent with the fundamental equation of dynamics 
 (\ref{qb1z}) from which it is deduced. We note, by contrast,
that the ordinary 
Schr\"odinger equation (\ref{qm2}) breaks the equivalence
principle since it
explicitly depends on the inertial mass of the particles \cite{nottale}. This
suggests that the Schr\"odinger-like
equation (\ref{qm1}) with a diffusion coefficient ${\cal D}$ may be more
relevant to describe astrophysical systems like  DM halos than the ordinary 
Schr\"odinger equation (\ref{qm2}).

{\it Remark:} For a free particle ($\Phi=0$) the Schr\"odinger-like
equation (\ref{qm1}) reduces to
\begin{equation}
\label{qm4}
\frac{\partial\psi}{\partial
t}=i{\cal D}\Delta\psi.
\end{equation}
Under this form, the Schr\"odinger-like equation is similar to a
diffusion equation with an imaginary diffusion coefficient $i{\cal D}$. This
strengthens the equivalence between the Schr\"odinger equation
and the complex
Burgers equation 
\begin{equation}
\label{qb2sw}
\frac{\partial {\bf U}}{\partial t}+({\bf U}\cdot \nabla){\bf U}=i {\cal
D}\Delta{\bf U}
\end{equation}
through the complex Cole-Hopf transformation (\ref{qb4}).

\subsection{Fluctuation-dissipation theorem}

We now come back to the generalized Schr\"odinger equation (\ref{qb6}) with
$\gamma\neq 0$ including dissipation.
Writing $\gamma=\gamma_R+i\gamma_I$, where $\gamma_R$ is the classical friction
coefficient and $\gamma_I$ is the
quantum friction coefficient, and using  the identity
\begin{equation}
\label{we1}
{\rm Im}(\gamma\ln\psi)=\gamma_I\ln|\psi|-\frac{1}{2}i \gamma_R\ln\left
(\frac{\psi}{\psi^*}\right ),
\end{equation}
we can rewrite Eq. (\ref{qb6}) as
\begin{eqnarray}
\label{we2}
i{\cal D}\frac{\partial\psi}{\partial t}=-{\cal D}^2\Delta\psi
+\frac{1}{2}\Phi\psi+\frac{1}{2}V(t)\psi\nonumber\\
+{\cal D} \gamma_I\ln|\psi|\,
\psi-i\frac{{\cal D}}{2}\gamma_R\ln\left
(\frac{\psi}{\psi^*}\right )\, \psi.
\end{eqnarray}
Introducing the notations
\begin{equation}
\label{we3}
\gamma_R=\xi,\qquad \gamma_I=\frac{k_B T}{{\cal
D}m},
\end{equation}
the generalized Schr\"odinger equation (\ref{qb6}) takes the form
\begin{eqnarray}
\label{we4}
i{\cal D}\frac{\partial\psi}{\partial
t}=-{\cal
D}^2\Delta\psi+\frac{1}{2}\Phi\psi+\frac{1}{2}V(t)\psi\nonumber\\
+\frac{k_B T}{m}\ln|\psi|\, \psi-\frac{1}{2}i\xi{\cal D}\ln\left
(\frac{\psi}{\psi^*}\right )\, \psi.
\end{eqnarray}
Using the hydrodynamic representation of the
generalized Schr\"odinger equation (see below), we can interpret  $\xi$ 
as an ordinary friction coefficient and $T$ as an
effective temperature ($k_B$ is Boltzmann's constant). It is
convenient to
choose
the function $V(t)$ so that the average value of the friction term proportional
to $\xi$ is equal to zero. This gives
\begin{equation}
\label{we5}
V(t)=i\xi{\cal D}\left\langle \ln\left (\frac{\psi}{\psi^*}\right
)\right\rangle,
\end{equation}
where $\langle X\rangle=(1/M)\int \rho X\, d{\bf r}$. We finally obtain the
generalized Schr\"odinger equation\footnote{It is possible to generalize this
equation further by taking the self-interaction of the particles into account
(see Appendices \ref{sec_sr} and \ref{sec_lb}).}
\begin{eqnarray}
\label{we6}
i{\cal D}\frac{\partial\psi}{\partial
t}=-{\cal D}^2\Delta\psi+\frac{1}{2}\Phi\psi+\frac{k_B
T}{m}\ln|\psi|\, \psi\nonumber\\
-\frac{1}{2}i\xi{\cal D}\left\lbrack \ln\left
(\frac{\psi}{\psi^*}\right )-\left\langle \ln\left (\frac{\psi}{\psi^*}\right
)\right\rangle\right\rbrack\, \psi.
\end{eqnarray}
This equation has to be coupled to the Poisson equation
(\ref{dw2}). They can be combined into a single differential
equation 
\begin{eqnarray}
\label{weqpsec}
i{\cal D}\frac{\partial\Delta\ln\psi}{\partial
t}=-{\cal
D}^2\Delta\left
(\frac{\Delta\psi}{\psi}\right )+2\pi G|\psi|^2\nonumber\\
+\frac{k_B
T}{m}\Delta\ln|\psi|-\frac{1}{2}i\xi{\cal D}\Delta\ln
\left(\frac{\psi}{\psi^*}\right ).
\end{eqnarray}

It is interesting to note that the complex nature of the
friction coefficient 
\begin{equation}
\label{we7}
\gamma=\xi+i\frac{k_B T}{{\cal D}m}
\end{equation}
leads to a generalized Schr\"odinger equation exhibiting {\it simultaneously} a
friction term (as expected) 
and an effective temperature term (unexpected). They correspond to the
real and imaginary parts of $\gamma$. This may be viewed as a new form of
fluctuation-dissipation theorem. In this respect, we note that
the relation
\begin{equation}
{\cal D}=\frac{k_B T}{m\gamma_I}
\end{equation}
is similar to the Einstein relation of
Brownian motion \cite{einstein}. It is important to stress,
however, that $T$ does not represent the true
thermodynamic temperature which is here assumed to be equal to zero (see
Appendix \ref{sec_tef}). It could be
interpreted as the temperature of an hypothetical Dirac-like
aether \cite{diracaether} (it may represent the
temperature of the vacuum if it has fluctuations), or be  an intrinsic
property of the fractal spacetime itself. Similarly, the
friction coefficient $\xi$ may characterize the friction of the system with the
aether or be an intrinsic property of the fractal spacetime.
Another possibility is that the effective temperature $T$ and
the friction $\xi$ heuristically parametrize the process of violent relaxation
and gravitational cooling experienced by the system on a coarse-grained scale.

{\it Remark:} In relation to the equivalence principle discussed in
Sec. \ref{sec_rec}, a
comment may be in
order. It seems that the mass $m$ of the particles has appeared in Eq.
(\ref{we6}). However, its occurence is artificial because it arises from the
notation of Eq. (\ref{we3}) that involves the effective temperature $T$. In
fact, only the
ratio $k_B T/m$  matters and this ratio is independent of the mass (in other
words, $m$ is an effective mass that needs not coincide with the DM particle
mass). By
anticipating a result that will be obtained below, we could have written
$v_\infty^2/2$ instead of $k_B T/m$ so that the mass $m$ does not appear
anymore in the equations.\footnote{This notation makes sense
since the coefficient in front of $\ln|\psi|\psi$ in Eq. (\ref{we4}) has the
dimension of a velocity square. As we shall see, the velocity  $v_\infty$
corresponds to the constant circular velocity of the spiral
galaxies.}  In this way, the equivalence principle is respected.
However, in order to develop an analogy with thermodynamics (see below), we
shall work in terms of an effective temperature $T$ and an effective mass $m$.

\subsection{The Madelung transformation}

Using the Madelung \cite{madelung} transformation, the generalized Schr\"odinger equation
(\ref{we6}) can be written  in the form of real hydrodynamic
equations. To that purpose, we write
the wave function as
\begin{equation}
\label{ntmad2}
\psi({\bf r},t)=\sqrt{\rho({\bf r},t)} e^{i\sigma({\bf r},t)/2{\cal D}},
\end{equation}
where $\rho$ is the density and $\sigma$ is a real potential or a real action.
They can be expressed in terms of the wave function as
\begin{equation}
\label{ntmad2b}
\rho=|\psi|^2,\qquad \sigma=-i{\cal D}\ln\left (\frac{\psi}{\psi^*}\right
).
\end{equation}
We note that the effective temperature term  in the generalized Schr\"odinger
equation
(\ref{we6}) can
be written as $(k_B T/2m)\ln\rho\, \psi$ and the dissipative term as
$(1/2)\xi(\sigma-\langle \sigma\rangle)\psi$. Following Madelung, we introduce
the real potential velocity field ${\bf
u}=\nabla\sigma$. The flow defined in this way is irrotational since
$\nabla\times {\bf u}={\bf
0}$. Substituting Eq. (\ref{ntmad2}) into the generalized Schr\"odinger equation
(\ref{we6}) and separating the real and imaginary parts, we
find that  the generalized Schr\"odinger equation is equivalent to
the
hydrodynamic equations
\begin{equation}
\label{tmad4}
\frac{\partial\rho}{\partial t}+\nabla\cdot (\rho {\bf u})=0,
\end{equation}
\begin{equation}
\label{tmad5}
\frac{\partial \sigma}{\partial t}+\frac{(\nabla
\sigma)^2}{2}+\Phi+V_Q+\frac{k_B T}{m}\ln\rho+\xi
(\sigma-\langle \sigma\rangle)=0,
\end{equation}
where
\begin{equation}
\label{ntmad8}
V_Q=-2{\cal D}^2\frac{\Delta \sqrt{\rho}}{\sqrt{\rho}}
\end{equation}
is the quantum potential. The first equation is the
continuity equation and the second equation is the
quantum Hamilton-Jacobi equation for a real action or the quantum Bernoulli
equation for a real potential. 
Taking the gradient of Eq. (\ref{tmad5}), we obtain the quantum damped
isothermal Euler equation
\begin{equation}
\label{tmad6}
\frac{\partial {\bf u}}{\partial t}+({\bf u}\cdot \nabla){\bf
u}=-\frac{1}{\rho}\nabla P-\nabla\Phi-\nabla V_Q-\xi {\bf u}.
\end{equation}
It involves a
pressure force with an effective isothermal equation of state 
\begin{equation}
\label{tmad7}
P=\rho \frac{k_B T}{m},
\end{equation}
a gravitational force, a quantum force, and a damping force. 
Using the continuity equation (\ref{tmad4}), the quantum damped isothermal Euler
equation (\ref{tmad6}) can can be rewritten as
\begin{equation}
\label{tmad9}
\frac{\partial}{\partial t}(\rho {\bf u})+\nabla(\rho {\bf u}\otimes {\bf u})
=-\nabla P-\rho\nabla\Phi-\rho\nabla V_Q-\xi\rho {\bf u}.
\end{equation}
When the quantum potential is neglected (Thomas-Fermi approximation), we recover
the classical damped
isothermal Euler equation. For dissipationless systems
($\xi=0$), we recover the quantum and classical isothermal
Euler equations.

\subsection{Connection with the equations of Brownian theory}
\label{sec_conn}

In the overdamped limit $\xi\rightarrow +\infty$, we can
formally neglect the inertial term in Eq. (\ref{tmad6}) so that
\begin{equation}
\label{tmad10}
\xi{\bf u}\simeq -\frac{1}{\rho}\nabla P-\nabla\Phi-\nabla V_Q.
\end{equation}
Substituting this relation into the continuity equation (\ref{tmad4}), we 
obtain the quantum  Smoluchowski  equation \cite{pre11}:
\begin{equation}
\label{tmad11}
\xi\frac{\partial\rho}{\partial t}=\nabla\cdot\left (\nabla
P+\rho\nabla\Phi+\rho\nabla V_Q\right ).
\end{equation}
When the quantum potential is neglected, it reduces to the
classical Smoluchowski
equation 
\begin{equation}
\label{tmad11b}
\xi\frac{\partial\rho}{\partial t}=\nabla\cdot\left (\frac{k_B T}{m}\nabla\rho
+\rho\nabla\Phi\right )
\end{equation}
that was introduced in the  context of
Brownian motion \cite{smoluchowski}. The diffusion coefficient
satisfies the standard Einstein relation  \cite{einstein}:
\begin{equation}
D=\frac{k_B T}{\xi m}.
\end{equation}
On the other hand, if we neglect the advection term $\nabla(\rho {\bf u}\otimes
{\bf u})$
in Eq. (\ref{tmad9}), but retain the term $\partial (\rho{\bf u})/\partial t$,
and combine the
resulting
equation with the continuity equation (\ref{tmad4}), we obtain the quantum
telegraph  equation
\begin{equation}
\label{tmad12}
\frac{\partial^2\rho}{\partial t^2}+\xi\frac{\partial\rho}{\partial
t}=\nabla\cdot\left (\nabla P+\rho\nabla\Phi+\rho\nabla V_Q\right )
\end{equation}
which can be seen as a generalization of the quantum Smoluchowski equation
(\ref{tmad11}) taking inertial (or memory) effects into account.
When the quantum potential is neglected, we recover the
classical telegraph equation.

It is interesting to recover the equations of Brownian
theory, with a completely different
interpretation, from the generalized Schr\"odinger
equation (\ref{we6}) in a strong
friction limit. In this sense, our approach
makes a connection between quantum mechanics and Brownian motion.
However, we emphasize
that (besides the presence of the quantum potential) this analogy is
essentially formal. For example, the
diffusion term in the Smoluchowski equation for Brownian particles arises from
stochastic processes (it is due to a random force or a noise in the Langevin
\cite{langevin}
equations of Brownian motion) while the
diffusion term in the Smoluchowski equation derived from the generalized
Schr\"odinger equation (\ref{we6}) arises from a logarithmic nonlinearity
stemming from the imaginary part of the complex friction coefficient in the
covariant
equation of dynamics (\ref{qb1}). One is left speculating if
this complex friction force is equivalent to a stochastic force. In any case, at
a formal level, the generalized Schr\"odinger equation (\ref{we6}),
which is equivalent 
to the covariant equation of dynamics (\ref{qb1}), unifies quantum
mechanics and Brownian motion.

{\it Remark:} The dynamics and thermodynamics of self-gravitating Brownian
particles described
by the Smoluchowski-Poisson equations  (\ref{poisson}) and (\ref{tmad11b}) has
been studied in \cite{sp1,sp2,sp3,sp4,sp5,sp6}. If the strong friction
limit $\xi\rightarrow +\infty$ is relevant, this work could be applied to the
generalized Schr\"odinger-Poisson equations (\ref{dw2}) and (\ref{we6}). 

\subsection{Justification of the Born interpretation}

 In the theory of scale relativity, the fundamental object of interest is the
complex velocity ${\bf U}$ of the fractal geodesics and the complex hydrodynamic
equation (\ref{qb2}) of these
geodesics from which the (generalized) Schr\"odinger equation (\ref{we6}) can be
derived. 
Now, we expect the fluid of geodesics to be more concentrated at some places and
less at others as does a real fluid. To find the probability density of presence
of the paths we can
remark that Eqs. (\ref{qb2}) and (\ref{we6})  are equivalent to the
real hydrodynamic
equations (\ref{tmad4})-(\ref{tmad9}). In that
context, ${\bf u}=\nabla\sigma$ is not an {\it ad hoc} definition (unlike in
Madelung's original work) but it corresponds to the classical velocity (the real
part of ${\bf U}$). On the other hand, in the
theory of scale relativity, the
quantum potential is a manifestation of the geometry of spacetime, namely, of
its non-differentiability and fractality, in similarity with  the Newtonian
potential being a manifestation of the curvature of spacetime in Einstein's
theory of general relativity \cite{nottale}. Then, Eqs.
(\ref{tmad4})-(\ref{tmad9}) describe a fluid of fractal geodesics in a
non-differentiable spacetime.
They have therefore a clear physical interpretation. As a result $\rho({\bf
r},t)=|\psi|^2$ represents the density of the geodesic fluid, and the
probability
density for the ``quantum particle'' to be found at a given position
must
be proportional to  $|\psi|^2$. This is how the
Born postulate can be naturally justified in the theory of
scale relativity \cite{nottale}.

 \section{Application to DM halos}
\label{sec_app}

We now apply the generalized  Schr\"odinger equation (\ref{we6}) to the context
of DM halos and show qualitatively how it can account for their main properties.
A more quantitative study, and a comparison with observations, will be the
subject of a specific paper \cite{prep}.

\subsection{Quantum hydrostatic equilibrium}

Considering a solution of 
the generalized Schr\"odinger  equation (\ref{we6}) of the form $\psi({\bf
r},t)=\phi({\bf
r})e^{-i\epsilon t/2{\cal D}}$  where $\phi({\bf
r})$ and $\epsilon$ (energy) are real, we obtain the time-independent
generalized
Schr\"odinger  equation
\begin{equation}
\label{app1}
-2{\cal D}^2\Delta\phi+\Phi\phi+2\frac{k_BT}{m}(\ln\phi)\phi=\epsilon\phi,
\end{equation}
which is a nonlinear eigenvalue equation. Dividing Eq.
(\ref{app1}) by $\phi$ and using the fact that $\phi({\bf
r})=\sqrt{\rho({\bf r})}$, or directly substituting $\sigma=-\epsilon t$ into
Eq. (\ref{tmad5}), we obtain the equilibrium
Hamilton-Jacobi (or Bernoulli) equation
\begin{equation}
\label{app2}
-2{\cal
D}^2\frac{\Delta\sqrt{\rho}}{\sqrt{\rho}}+\Phi+\frac{k_BT}{m}\ln\rho=\epsilon.
\end{equation}
Taking the  gradient of Eq. (\ref{app2}), we obtain the condition of
quantum hydrostatic
equilibrium
\begin{eqnarray}
\label{app3}
\rho\nabla V_Q+\nabla P+\rho\nabla\Phi={\bf 0}.
\end{eqnarray}
This equation corresponds also to the equilibrium state (${\bf
u}={\bf 0}$) of the quantum Euler equation
(\ref{tmad6}).
It describes the balance between the gravitational attraction, the quantum
pressure, and the effective thermal pressure. It must be coupled
self-consistently to the Poisson
equation (\ref{poisson}). Equation (\ref{app2}) can be rewritten as
\begin{equation}
\label{app3b}
\rho=Ae^{-\frac{m}{k_B T}(\Phi+V_Q)} \quad {\rm with}\quad A=e^{m\epsilon/k_B
T},
\end{equation}
which can be interpreted as a quantum Boltzmann distribution. This is actually
a differential equation since the quantum potential $V_Q$ involves derivatives
of the density.

\subsection{$H$-theorem}
\label{sec_h}

Introducing the Boltzmann free energy $F_B=E-TS_B$
where $E=\Theta_c+\Theta_Q+W$ is the energy (including the classical
kinetic energy $\Theta_c=(1/2)\int\rho {\bf
u}^2\, d{\bf r}$, the quantum kinetic energy $\Theta_Q=\int \rho
V_{Q}\, d{\bf
r}$ and the gravitational energy $W=({1}/{2})\int\rho\Phi\, d{\bf r}$) and
$S_B=-k_B \int (\rho/m)(\ln\rho-1)\,
d{\bf r}$ is the Boltzmann entropy, we can show that the generalized
Schr\"odinger 
equation (\ref{we6}) satisfies an $H$-theorem \cite{total}:
\begin{eqnarray}
\label{app4}
\dot F_B=-\xi\int \rho {\bf u}^2\, d{\bf r}\le 0.
\end{eqnarray}
When $\xi=0$, the generalized Schr\"odinger equation (\ref{we6}) conserves
the energy ($\dot F_B=0$). When $\xi>0$, the free energy decreases
monotonically ($\dot F_B\le 0$). On the other hand, $\dot F_B=0$ if, and
only if, ${\bf
u}={\bf 0}$, leading to the condition of quantum hydrostatic equilibrium
(\ref{app3}). Therefore, the dissipative term ensures that the
system relaxes
towards an equilibrium state  for $t\rightarrow +\infty$.\footnote{This result
assumes that $F_B$ is bounded from below. For isothermal self-gravitating
systems this is not the case. There is no minimum of free energy at fixed mass
because the system can always loose free energy by evaporating except if it is
artificially confined within a box \cite{paddy}. However, evaporation is
a slow process. In practice, the system relaxes towards a quasiequilibrium
state that slowly evolves in time because of
evaporation.}
In this sense, it can account for the complicated processes of violent
relaxation and gravitational cooling. The 
equilibrium state minimizes the Boltzmann free energy  $F_B$ at fixed mass
$M$ (see footnote 13).
Writing the
variational principle as $\delta F-\alpha\delta M=0$, where $\alpha$ is a
Lagrange multiplier (chemical potential)  taking
into account the conservation of
mass, we recover Eq. (\ref{app2}) with $\alpha=\epsilon$. Therefore, the
eigenenergy
$\epsilon$ represents the chemical potential $\alpha$.

{\it Remark:} We note that the conservative ($\xi=0$)
generalized Schr\"odinger equation (\ref{we6}) is reversible while the
dissipative ($\xi>0$) generalized Schr\"odinger equation (\ref{we6}) is
irreversible. As discussed previously, irreversibility may be the manifestation
of a cosmic aether or the consequence of a process of violent relaxation and
gravitational cooling on a coarse-grained scale (a sort of nonlinear Landau
damping like for the Vlasov-Poisson equations \cite{villani}).

\subsection{General differential equation}

Combining the equation of hydrostatic equilibrium (\ref{app3}) and the Poisson
equation (\ref{poisson}), and using the effective isothermal equation of
state (\ref{tmad7}), we obtain a differential equation
 \begin{eqnarray}
\label{app5}
2{\cal D}^2\Delta
\left (\frac{\Delta\sqrt{\rho}}{\sqrt{\rho}}\right
)-\frac{k_B
T}{m}\Delta\ln\rho
=4\pi
G\rho
\end{eqnarray}
that determines the density profile of the DM halos. This
profile has a
core-halo
structure that is studied in detail in a separate paper \cite{prep} where Eq.
(\ref{app5})
is solved numerically. We describe below
how Eq. (\ref{app5}) simplifies in the core and in the halo respectively.
Then, we give a preliminary discussion of its general solution 
and show how
it can account for the main properties of DM halos.

\subsubsection{Solitonic core}

In the core, thermal effects are
negligible and the condition
of hydrostatic equilibrium (\ref{app3}) reduces to
\begin{equation}
\label{app6}
\nabla V_Q+\nabla\Phi=0.
\end{equation}
It describes the balance between the gravitational attraction and the quantum
pressure arising from the Heisenberg uncertainty principle. The differential
equation (\ref{app5}) becomes
 \begin{eqnarray}
\label{app7}
2{\cal D}^2\Delta
\left (\frac{\Delta\sqrt{\rho}}{\sqrt{\rho}}\right
)=4\pi
G\rho.
\end{eqnarray}
This equation has been solved numerically in
\cite{rb,membrado,gul0,gul,prd2,ch2,ch3,pop}.
 Its
solution is called a soliton because it
corresponds to the static state of the ordinary Schr\"odinger-Poisson  equation.
This profile presents a core in which the central density is
finite. As
a result, it can solve the cusp
problem of CDM.  The exact mass-radius relation is given by
\cite{rb,membrado,prd2}:
\begin{equation}
\label{app8}
R_{99}=39.6 \frac{{\cal D}^2}{GM},
\end{equation}
where $R_{99}$ is the radius of the configuration containing $99\%$
of the mass. The density profile extends to
infinity. It has an approximately Gaussian shape \cite{prd1,prd2}. 
Another fit of this profile has been given in
\cite{ch2,ch3}.

\subsubsection{Isothermal halo}

In the halo, quantum effects are negligible and the condition
of hydrostatic equilibrium (\ref{app3}) reduces to
\begin{equation}
\label{app9}
\nabla P+\rho\nabla\Phi=0.
\end{equation}
It describes the balance between the gravitational attraction and the
effective thermal pressure. Using Eq. (\ref{tmad7}), it can be integrated
into
\begin{equation}
\label{app10}
\rho=Ae^{-m\Phi/k_B T},
\end{equation}
which can be interpreted as Boltzmann's law in a gravitational mean
field potential. This equation must be coupled self-consistently to the
Poisson equation (\ref{poisson}) leading to the Boltzmann-Poisson equation
\begin{equation}
\label{app11}
\Delta\Phi=4\pi G Ae^{-m\Phi/k_B T}.
\end{equation}
This equation arises in the statistical mechanics of
self-gravitating systems \cite{paddy} but it has been derived here from
rather different arguments.  Equation (\ref{app11}) is equivalent to the
differential
equation
\begin{equation}
\label{app12}
\Delta \ln\rho+\frac{4\pi Gm}{k_B T}\rho=0
\end{equation}
obtained from Eq. (\ref{app5}) by neglecting the quantum potential.
It is easy to show that the asymptotic behavior of the density distribution is
given by \cite{bt}:
\begin{equation}
\label{app13}
\rho(r)\sim_{+\infty}\frac{k_B T}{2\pi Gm r^2}.
\end{equation}
The mass contained within a sphere of radius $r$ behaves at large distances as
$M(r)=\int_0^r \rho(r') 4\pi {r'}^2\, dr'\sim 2k_B
Tr/Gm$.\footnote{We note that the mass of a self-gravitating
isothermal sphere diverges as $r\rightarrow +\infty$. This is the
so-called infinite mass problem \cite{bt}. This is why there is no minimum of
free energy at fixed mass in an unbounded
domain. If we want to have a true equilibrium state, we must confine the system
within a box \cite{paddy}. However, in practice, we
are not interested by the behavior of the profile at infinity because 
physical processes will confine the system within a finite region of space.
Therefore, an isothermal profile may be relevant to describe DM halos over the
range of distances corresponding to the observations.} Therefore, the
(effective)
isothermal density profile
leads to flat rotation curves  since \cite{bt}:
\begin{equation}
\label{app14}
v_{c}^2(r)=\frac{GM(r)}{r}\rightarrow \frac{2k_B T}{m} \quad {\rm for}\quad
r\rightarrow +\infty,
\end{equation}
where $v_c(r)$ is the circular velocity at distance $r$. This
result is consistent
with the observations that show that the  circular velocity of spiral galaxies
tends to a constant $v_{\infty}$ at large distances instead of declining
according to Kepler's law. We
find that $v_{\infty}=(2k_B T/m)^{1/2}$. We stress again that the effective
temperature $T$ does not correspond to the thermodynamic
temperature which is zero in
the situation that we consider here ($T_{\rm thermo}= 0$).
Still, observations reveal
that DM halos have an almost ``isothermal'' atmosphere yielding flat rotation
curves. In this paper, we suggest that this effective isothermal
atmosphere could be the cosmological
manifestation of the logarithmic nonlinearity $(k_B
T/m)\ln|\psi|$, or $(v_\infty^2/2)\ln|\psi|$, present in
the generalized Schr\"odinger equation (\ref{we6}). As
mentioned previously, it could be regarded as (i) the temperature of an
hypothetical
aether (or the temperature of the vacuum), (ii) an intrinsic property of the
fractal spacetime itself, or (iii)  a consequence of violent relaxation and
gravitational cooling on a
coarse-grained scale.

We finally note that observations
\cite{observations} and numerical simulations
\cite{nfw,ch2,ch3}  show that the atmosphere of DM halos is closer to the NFW
and Burkert profiles (decaying at large distances as $r^{-3}$) than to the
isothermal
profile (decaying as $r^{-2}$). This difference may be due to
complicated physical effects such as incomplete relaxation, tidal
effects, or external stochastic forcing.\footnote{From general
thermodynamical arguments, we expect the system to reach an isothermal
distribution. However, in practice, nonideal effects may prevent its
relaxation towards thermodynamical equilibrium. This
is relatively obvious in the present context since a self-gravitating
isothermal system has an infinite mass \cite{bt}. Therefore, its atmosphere
cannot be exactly isothermal. We note in this respect that
the exponent $\alpha=-3$ (NFW/Burkert) of the observed density profile
$\rho\sim r^{-\alpha}$ at large distances is the closest
exponent to $\alpha=2$ (isothermal) that
yields a halo with a (marginally) finite
mass.} The complicated problem of incomplete relaxation has been addressed in
the context of
Lynden-Bell's theory of violent relaxation \cite{lb} and similar arguments
could be developed here to explain the deviation between the isothermal
profile and the NFW/Burkert profiles at large distances. Another
possibility would
be to heuristically introduce a confining potential in the generalized
Schr\"odinger
equation (\ref{we6}) to steepen the density profile. However, in a first
approximation, an isothermal
envelope provides a fair description of DM halos and has the nice feature to
yield exactly flat rotation curves.

\subsubsection{Core-halo structure}
\label{sec_gde}

\begin{figure}
\begin{center}
\includegraphics[clip,scale=0.3]{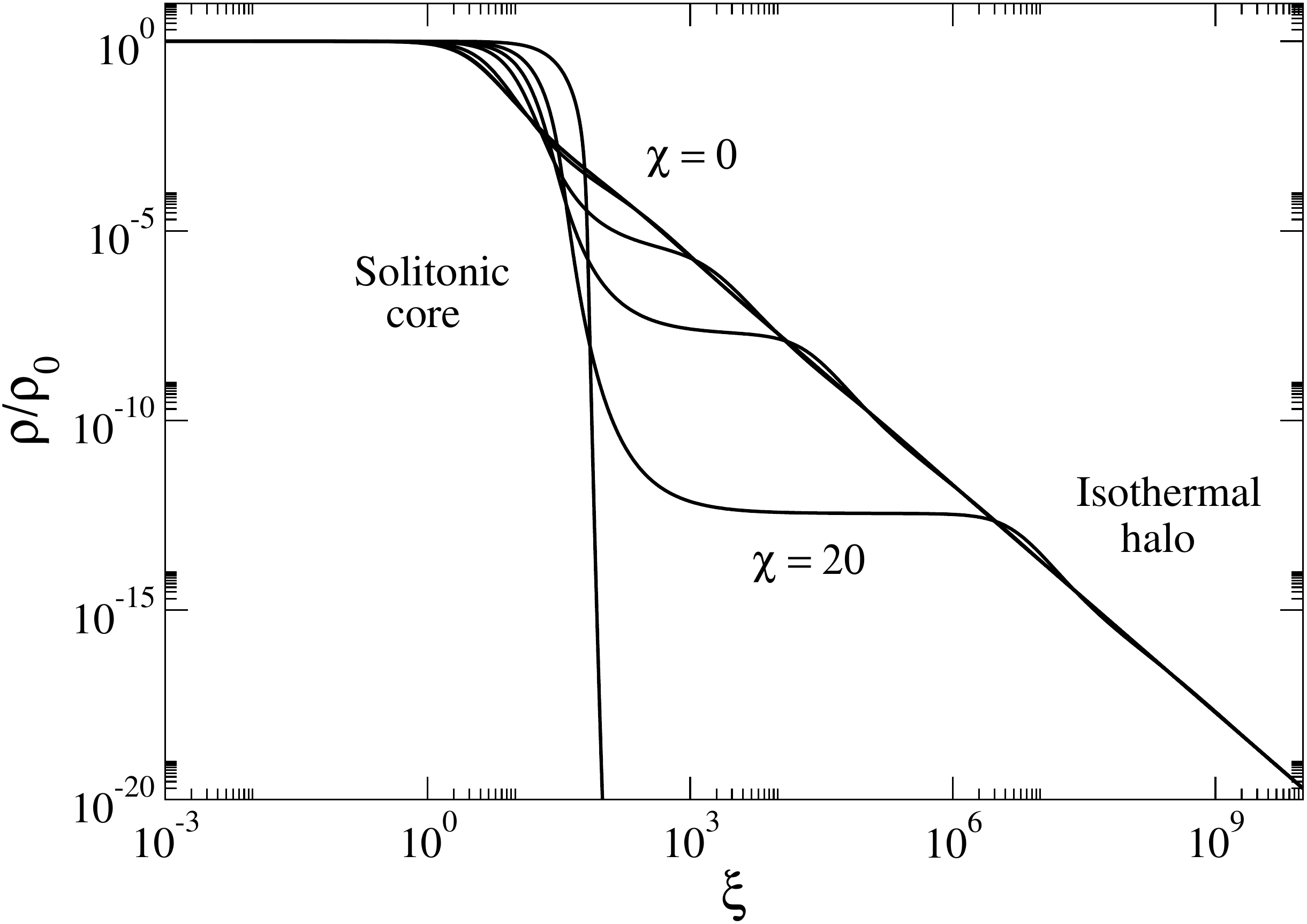}
\caption{Normalized density  profiles for different values of
the concentration
parameter $\chi$ (specifically $\chi=0,0.1,1,5,10,20,100$). They present a
core-halo structure with a solitonic core and an isothermal halo. The purely
isothermal halo corresponds to $\chi=0$ and the purely  solitonic
profile corresponds to $\chi\rightarrow +\infty$.}
\label{densite}
\end{center}
\end{figure}

\begin{figure}
\begin{center}
\includegraphics[clip,scale=0.3]{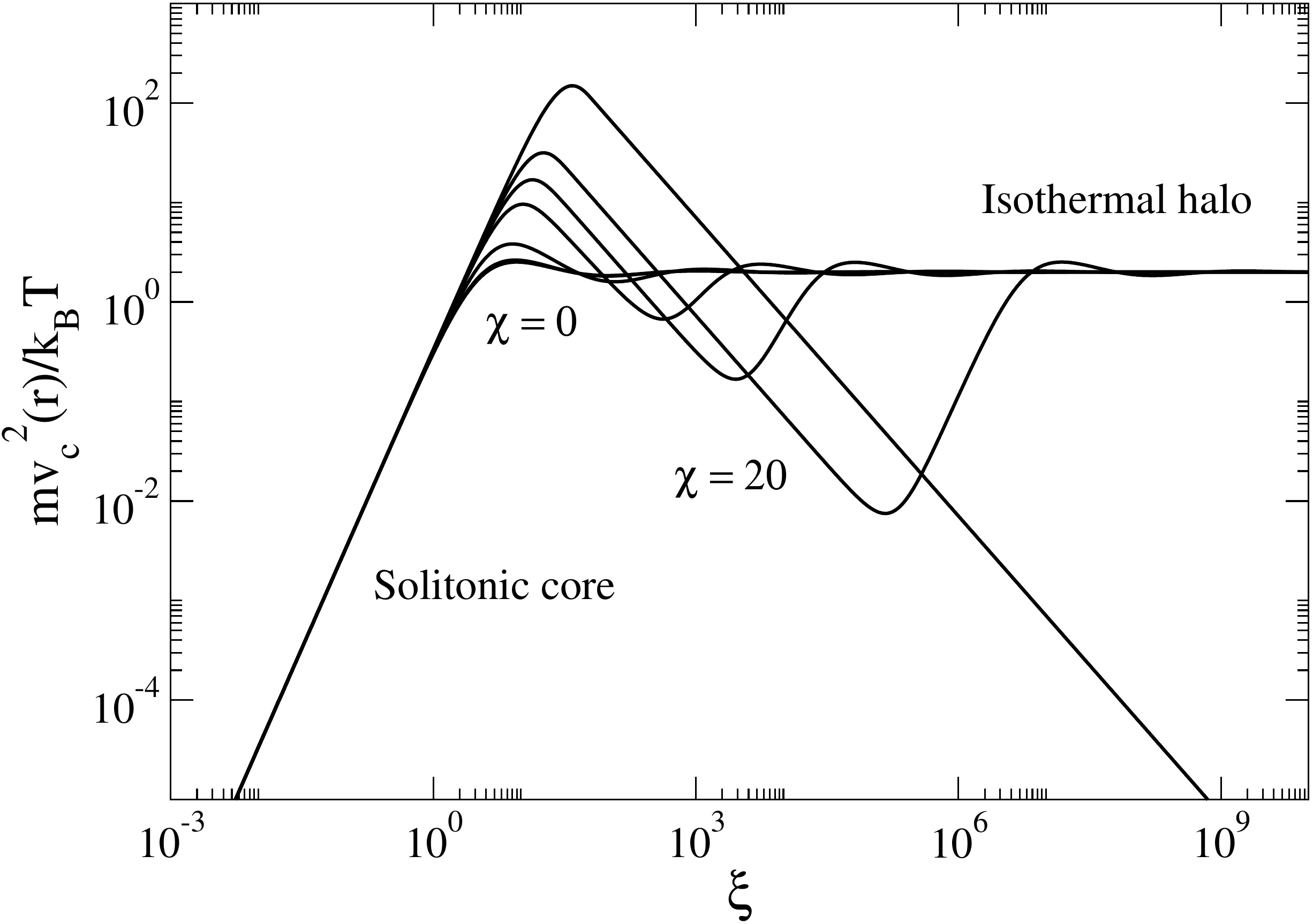}
\caption{Same as Fig. \ref{densite} for the circular velocity
profiles. They
display a dip due to the presence of the solitonic
core.}
\label{vitesse}
\end{center}
\end{figure}

In this section, we give a preliminary discussion of the
differential equation (\ref{app5}) governing the structure of DM halos in the
present model. If we define
\begin{equation}
\label{gde2}
\rho=\rho_0
e^{-\psi},\qquad  \xi=\left (\frac{4\pi G\rho_0
m}{k_B T}\right )^{1/2}r, 
\end{equation}
\begin{equation}
\label{gde3}
\chi=\frac{{\cal D}m}{k_B T}\sqrt{2\pi
G\rho_0},
\end{equation}
where $\rho_0$ is the central density and $\chi$ is a dimensionless
concentration parameter, we find that Eq.
(\ref{app5}) takes the form of a quantum Emden equation 
\begin{equation}
\label{gde4}
4\chi^2 \Delta\left (\frac{\Delta
e^{-\psi/2}}{e^{-\psi/2}}\right )+\Delta\psi=e^{-\psi}.
\end{equation}
The classical Emden equation is recovered for $\chi=0$ \cite{chandra}.  The
differential equation (\ref{gde4}) can be solved numerically. The density
profiles and the circular velocity profiles are plotted in Figs. \ref{densite}
and \ref{vitesse} for different values of $\chi$. The density profile presents a
core-halo
structure which is more or less pronounced depending on the value of
the concentration parameter $\chi$. The velocity profile shows a dip which is
due to the presence of the solitonic core. These profiles are qualitatively
similar to the core-halo profiles obtained by Schive {\it et al.}
\cite{ch2,ch3} by solving the Schr\"odinger-Poisson equations numerically but
they are
not exactly the same. In particular, in the present model, the halo is
isothermal instead
of being given by the NFW profile. It will be of interest in future works to
compare these profiles with real DM halos. The present paper is just a first
step in that direction. A more detailed study will be performed in a
separate
paper \cite{prep}.

\subsection{The fundamental parameters of the generalized Schr\"odinger
equation}
\label{sec_fund}

\subsubsection{The coefficient ${\cal D}$}
\label{sec_coeffD}

As discussed previously, the equilibrium states of the generalized
Schr\"odinger
equation (\ref{we6}) have a core-halo structure that is in qualitative
agreement with the structure of DM halos \cite{ch2,ch3}. The quantumlike
potential accounts for their solitonic core (solving the cusp problem) and the
logarithmic nonlinearity accounts for their isothermal halo (leading to flat
rotation curves). The extension of the atmosphere, as compared to that of the
core, depends on the size of the halos through the
concentration parameter $\chi$. Large DM halos such as
the Medium Spiral ($R\sim 10^4\, {\rm pc}$ and $M\sim 10^{11}\, M_\odot$) have a
core-halo structure with a small core and a large
atmosphere.\footnote{Actually, in large DM halos, the solitonic
core is almost imperceptible. In that case, the presence of a core (instead of a
cusp) is due to the effective thermal pressure rather than the quantum
potential.}
By contrast, small
DM halos such as dwarf spheroidal
galaxies (dSph) like Fornax ($R\sim 1\, {\rm kpc}$ and $M\sim 10^8\,
M_\odot$) are very compact and do not have an atmosphere. They are purely
solitonic, corresponding to the ground
state of the Schr\"odinger-Poisson equations. Therefore, their mass-radius
relation is given by Eq. (\ref{app8}). We shall determine the coefficient ${\cal
D}$ arising in the generalized Schr\"odinger equation (\ref{we6}) by
considering a typical dwarf halo of radius $R=1\, {\rm kpc}$ and mass
$M=10^8\,
M_\odot$ (Fornax).   Assuming that this halo represents the ground state of the
Schr\"odinger-Poisson equation (pure soliton) we find from Eq. (\ref{app8}) that
\begin{equation}
\label{d}
{\cal D}=1.02\times 10^{23}\, {\rm m^2/s}.
\end{equation}
If we assume that the nondifferentiability of spacetime is due to ordinary
quantum mechanics, using Eq. (\ref{qm3}) we find that the mass of the bosons
must be ultralight, of the order of $m=2.92\times 10^{-22}\,
{\rm eV}/c^2$ (in comparison, for the
electron of mass
$m=511\, {\rm
keV/c^2}$, the quantum diffusion coefficient ${\cal D}=\hbar/2m=5.79\times
10^{-5}\, {\rm m^2/s}$). However, the message conveyed in this paper is that 
the nondifferentiability of spacetime may arise from reasons different from
ordinary quantum mechanics (as explained in the Introduction). In that case, the
mass of the DM particle is not constrained (only ${\cal D}$ is fixed), allowing
for a larger class of particles including WIMPs, the QCD axion etc. The price to
pay with this new approach is that we have to take into account a new physical
ingredient in the equations of the problem, namely the fractal
structure of spacetime at the cosmic scale. This leads to a
complete reconsideration of the notion of space and time. This possibility is
not ruled out by observations since it is not possible to determine if the
Schr\"odinger equation that applies to DM halos is justified by the presence of
ultralight particles
(ordinary quantum mechanics) of by
the fractal structure of spacetime.

\subsubsection{The coefficients $v_\infty^2/2=k_B T/m$ and
$\xi$}
\label{sec_coeffT}

To determine the other coefficients $k_B T/m$ and $\xi$ that appear in the
generalized Schr\"odinger equation (\ref{we6}), we proceed as follows. It is
an
observational evidence that the surface density of DM halos is independent of
their size and has a  universal value $\Sigma_0=\rho_0r_h=141\, M_{\odot}/{\rm
pc}^2$ \cite{donato} where $\rho_0$ is the central density and $r_h$ is the halo
radius where the central density is divided by $4$. If we approximate large DM
halos by an isothermal sphere, one can show that $M_h=1.76 \Sigma_0 r_h^2$,
$v_h^2=1.76 G\Sigma_0 r_h$ and
$k_B T/m=0.954 G\Sigma_0r_h$ \cite{prep}. Therefore, the halo mass
scales
with size as $M_h\propto r_h^2$, the circular velocity at the
halo radius scales as $v_h\propto r_h^{1/2}$, and the temperature scales as $k_B
T/m\propto r_h$.
The dynamical time $t_{D}=1/(G\rho_0)^{1/2}=
(r_h/G\Sigma_0)^{1/2}$  scales as  $t_D\propto
r_h^{1/2}$. For a halo of mass $M_h=10^{11}\, M_\odot$ (Medium
Spiral), we find
$r_h=2.01\times 10^{4}\, {\rm pc}$, $\rho_0=7.02\times 10^{-3}M_{\odot}/{\rm
pc}^3$, $(k_B T/m)^{1/2}=108\, {\rm km/s}$, and  $t_{D}=178\, {\rm
Myrs}$ (we also have $v_h=(GM_h/r_h)^{1/2}=146\, {\rm km/s}$ and
$v_\infty=153\, {\rm km/s}$). Let us
write the friction coefficient as $\xi=\alpha t_D^{-1}$ where $\alpha$ is a
dimensionless parameter. The  friction coefficient is of the order of the
inverse relaxation time ($\xi\sim 1/t_R$). For self-gravitating systems
evolving as a result of two-body encounters, relaxation (thermalization) occurs
on a very long
timescale, of the order $t_R\sim (N/\ln N)t_D$ (Chandrasekhar's time)
\cite{bt}, much larger than the dynamical time.
In the situation that we consider here, an effective thermalization
may take place on a much shorter timescale, of the order of the dynamical time
$t_D$ (as in the process of violent relaxation of collisionless
self-gravitating systems
\cite{lb}), or even
shorter.
In order to take into account all the possibilities, we leave $\alpha$
arbitrary. As a result, the coefficients $k_B T/m=v_{\infty}^2/2$ and $\xi$ of
the generalized Schr\"odinger
equation (\ref{we6})
depend on the size of DM halos according to\footnote{The expression of $T$
is valid for relatively
large halos. Close to the ground state $T\rightarrow 0$.}
\begin{equation}
\label{dq}
\frac{v_{\infty}^2}{2}=\frac{k_B T}{m}=0.954 G\Sigma_0r_h,\qquad \xi=\alpha
\left (\frac{G\Sigma_0}{r_h}\right )^{1/2}.
\end{equation}
The scale dependence of $k_B T/m$ and $\xi$ is not a problem if the generalized
Schr\"odinger
equation (\ref{we6}) arises from the  fractal structure of spacetime at the
cosmological scale. In that case, its coefficient ${\cal D}$, $\xi$ and $k_BT/m$
are intrinsic properties of the spacetime  and they can change
with the size of the halos.\footnote{We have treated
the diffusion
coefficient ${\cal D}$ as a universal constant given by Eq. (\ref{d}).
The
diffusion coefficient ${\cal D}$  could also depend on the scale but
that
would add an indetermination in our
model that is not necessary (this
assumption of universality should be relaxed only if one finds that it is in
conflict with observations). Of course, the value of ${\cal D}$ given by Eq.
(\ref{d})
only applies to cosmological scales. It is very likely that ${\cal D}$ 
changes at smaller scales (galactic, stellar, planetary...). In this
respect, we note that the order of magnitude of ${\cal D}$ found by Nottale
\cite{nottale}
in the solar system is  ${\cal D}\sim 5\times 10^{14}\, {\rm
m^2/s}$.} The scale
dependence of $k_B T/m$ and $\xi$ is also expected if the effective temperature
and the friction result from a process of violent relaxation and gravitational
cooling. In that case, they will change from halo to halo depending on the
efficiency of the relaxation process.

{\it Remark:} If we assume that $\gamma_R\sim \gamma_I$
(since these coefficients have a common origin) and use
Eq. (\ref{we3}), we get ${\cal D}\sim D=k_B T/\xi m$. Using Eq. (\ref{dq}), we
obtain $\alpha\sim 0.954 (G\Sigma_0r_h^3)^{1/2}/{\cal D}$. For a halo of mass
$M_h=10^{11}\, M_\odot$ and radius $r_h=2.01\times 10^{4}\, {\rm
pc}$ (Medium
Spiral), we find $\alpha=\xi t_D\sim 640\gg 1$. This suggests  that large DM
halos are in the strong friction limit, allowing us to use the results of
\cite{sp1,sp2,sp3,sp4,sp5,sp6} valid for the Smoluchowski-Poisson
equations. However, the
assumption that
$\gamma_R\sim \gamma_I$ remains to be established on more solid grounds.
We give an argument in its favor in Appendix \ref{sec_einstein}
by generalizing to the case of dissipative systems the original method of
quantization introduced by Schr\"odinger \cite{schrodinger1}.

 \section{Conclusion}
\label{sec_conc}

In this paper, we have proposed to describe DM halos by the generalized
Schr\"odinger equation (\ref{we6}) obtained from the theory of scale
relativity \cite{nottale}. We have suggested that the origin of this equation
is not due to
ordinary quantum mechanics, as in the standard BECDM
model \cite{baldeschi,khlopov,membrado,sin,jisin,leekoh,schunckpreprint,
matosguzman,sahni,
guzmanmatos,hu,peebles,goodman,mu,arbey1,silverman1,matosall,silverman,
lesgourgues,arbey,fm1,bohmer,fm2,bmn,fm3,sikivie,mvm,lee09,ch1,lee,prd1,prd2,
prd3,briscese,
harkocosmo,harko,abrilMNRAS,aacosmo,velten,pires,park,rmbec,rindler,lora2,
abrilJCAP,mhh,lensing,glgr1,ch2,ch3,shapiro,bettoni,lora,mlbec,madarassy,
abrilph,playa,stiff,guth,souza,freitas,alexandre,schroven,pop,eby,cembranos,
braaten,davidson,schwabe,fan,calabrese,marsh,bectcoll,cotner,chavmatos,hui,
tkachevprl,abrilphas,shapironew,total,moczetal,phi6,
abriljeans,zhang,moczchavanis,desjacques}, but to the fractal
structure of spacetime (nondifferentiability) that manifests itself (i) above a
certain length scale (ii) and/or
for sufficiently long times. In the first case, this fractal structure is
regarded as an intrinsic property of spacetime at the astrophysical
or cosmological macroscale. In the second
case, it arises from the chaoticity of the trajectories of the 
particles.\footnote{According to Nottale's theory \cite{nottale}, a classical
system having a chaotic behavior may be
described by a Schr\"odinger-like equation on a very long
timescale.} As a result,
the coefficient ${\cal D}$ that arises
in the generalized Schr\"odinger equation (\ref{we6}) may be different from
$\hbar/2m$
allowing the mass of the DM particle to be much larger than the value $\sim
10^{-22}\, {\rm eV/c^2}$, corresponding to ultralight axions, required by the
BECDM model. Indeed, the observations of dwarf DM halos
(like Fornax) only determine the value of the coefficient ${\cal D}=1.02\times
10^{23}\, {\rm m^2/s}$, not
the mass $m$ of the DM particle. This is a consequence of the
equivalence principle. We have proposed that the value
of the coefficient ${\cal D}$ is universal at the
cosmological scale. It may, however, take a different value at the
astrophysical and planetary scales \cite{nottale}. By contrast, even at the
cosmological scale, the friction
parameter $\xi$ and the effective temperature $T$ are scale-dependent.
They behave
with the halos radius $r_h$ as $k_B T/m=0.954\, G\Sigma_0r_h$ and $\xi=640\,
({G\Sigma_0}/{r_h})^{1/2}$, where $\Sigma_0=141\, M_{\odot}/{\rm
pc}^2$ is the universal surface density of DM halos.

When applied to DM halos, 
the generalized Schr\"odinger equation (\ref{we6})  has attractive properties.
It
leads to equilibrium configurations with a core-halo structure similar to the
structure of DM halos observed in the Universe or in numerical simulations
\cite{ch2,ch3}. The quantum potential accounts for their solitonic core, the
logarithmic nonlinearity accounts for their isothermal halo yielding flat
rotation curves (it also accounts for the isothermal core of large DM halos),
and the friction term guarantees that the system relaxes
towards these core-halo structures (equilibrium states). This damping can
account for the
process of violent relaxation \cite{lb} and gravitational cooling
\cite{seidel94} on a coarse-grained scale. An interest of the
present  model is that there is no free (arbitrary) parameter. The coefficients
${\cal D}$, $k_B T/m$ and $\xi$ of the generalized Schr\"odinger equation
(\ref{we6}) are determined by Eqs. (\ref{d}) and (\ref{dq}). Therefore, for a
given halo mass $M_h$ (or radius $r_h$), one can predict the DM halo
profile by numerically solving
the
general differential equation (\ref{app5}). The results will be presented in
detail in a
forthcoming paper \cite{prep}. Preliminary results are given in \cite{total} and
in Figs. \ref{densite} and \ref{vitesse} of the present paper.

Apart from the reinterpretation of the coefficient ${\cal D}$, our approach is
similar to the BECDM model in which all the bosonic particles are in the same
quantum state described by the condensate wave function $\psi$.  However, the
difference of interpretation is crucial because, if correct, it would lead to
a complete reconsideration of the nature of space and time. In particular,
following Nottale \cite{nottale}, we suggest that spacetime could become fractal
(nondifferentiable) at the cosmic scales leading to a new form of
quantum mechanics. We stress that we do not reject the BECDM model 
based on ordinary quantum mechanics. This is at this day the most plausible
scenario. The purpose of this paper is just to propose an alternative  scenario
that has similar properties and can therefore solve the CDM crisis. It
could become particularly
interesting if one
finds that the mass of the DM particle is different from what is predicted by
the BECDM model.\footnote{In addition, our scenario has more
flexibility than ordinary quantum mechanics. For example, in our interpretation,
the
value of ${\cal D}$ could change with the scale if necessary (see footnote 18)
while this is not possible for ${\cal D}=\hbar/2m$ in quantum mechanics for
a given particle mass.}
Indeed, if
some arguments exclude ultralight particles, the existence of DM cores
(instead of cusps) could be 
a manifestation of the fractal (nondifferentiability) structure of spacetime at
the galactic
scale. To be very general, and
embrace all possibilities, we could consider a generalized Schr\"odinger
equation with a coefficient of the form ${\cal D}={\hbar}/{2m}+{\cal D}_{\rm
cosmo}$ where the first term is the contribution of standard quantum mechanics
(BECDM model) and the second term corresponds to the intrinsic fractal
structure of spacetime at the
cosmic scale.

Another interest of our formalism is to obtain general equations that, in a
sense, unify quantum
mechanics and Brownian theory (even if this unification is formal or 
effective). Indeed, in the
strong friction limit
$\xi\rightarrow +\infty$, the generalized Schr\"odinger equation (\ref{we6}) 
becomes equivalent to the quantum Smoluchowski equation (\ref{tmad11})
that is similar to the
one introduced  in Brownian theory (up to an additional quantum
potential).\footnote{At a formal level, the generalized
Schr\"odinger equation (\ref{we6}) decribes quantum motion ($\hbar\neq 0$ and
$\xi=0$), classical Brownian motion ($\hbar\rightarrow 0$ and
$\xi\rightarrow +\infty$), and quantum Brownian motion ($\hbar\neq 0$ and
$\xi\rightarrow +\infty$).  It is interesting to note that this (formal)
unification of quantum mechanics and Brownian theory is encapsulated in the
simple-looking
damped Newton equation (\ref{qb1}) that is equivalent to the generalized
Schr\"odinger 
equation (\ref{we6}). The imaginary part of the complex friction coefficient
plays the role of the stochastic force in the Langevin \cite{langevin} equations
of Brownian
theory. One can wonder whether this mysterious complex friction force is
equivalent to a stochastic force.} 
Self-gravitating Brownian particles described
by the 
Smoluchowski-Poisson
equations have been studied theoretically in
\cite{sp1,sp2,sp3,sp4,sp5,sp6}. Because of
this
formal analogy, if the strong friction limit is relevant to the case of DM (see
the Remark at the end of Sec. \ref{sec_coeffT}),   these theoretical
results on the dynamics
and thermodynamics of self-gravitating Brownian particles could find
applications in the physics of DM halos with a new
interpretation. This possibility will be considered in future works.

Finally, we would like to contrast the evolution of a
collisionless self-gravitating system evolving in a differentiable spacetime
with the evolution of a collisionless self-gravitating system
evolving in a nondifferentiable (fractal) spacetime. In the first case, the
system is described by the Vlasov-Poisson equations. These equations have a very
complicated dynamics associated with phase mixing and nonlinear Landau damping.
As a result, they develop intermingled filaments at smaller and smaller scales
and a coarse-grained description becomes necessary to smooth out this intricate
filamentation. The system is expected to relax
towards a quasistationary state (on the coarse-grained scale) on a very short
timescale of the order of the dynamical time. This is called violent
relaxation. Lynden-Bell \cite{lb} developed a statistical theory of this
process and derived an equilibrium distribution function describing this
quasistationary state. A sort of exclusion principle similar to the Pauli
exclusion principle in quantum mechanics arises in the theory of Lynden-Bell due
to the incompressibility of the flow in phase space and the conservation of the
distribution function (on the fine-grained scale) by the Vlasov equation. As a
result, the Lynden-Bell distribution function is similar to the Fermi-Dirac
distribution function, suggesting that the process
of violent relaxation is similar in some respects to the relaxation of fermionic
particles. A kinetic equation for the coarse-grained distribution function has
been proposed in \cite{csr,cg,dubrovnik}. It can be viewed as a generalized
Landau
or
Fokker-Planck equation taking into into account the Lynden-Bell exclusion
principle. We can then derive hydrodynamic equations \cite{csr} which have
the form of damped Euler equations including a pressure term with a
fermionic-like equation of state. These hydrodynamic equations do not
involve a quantum potential because the analogy
between the Lynden-Bell theory and quantum mechanics is purely
effective.\footnote{The Lynden-Bell theory accounts for an exclusion principle
arising from the Vlasov equation which plays a role similar to the Pauli
exclusion principle in quantum mechanics. This is why the Lynden-Bell
distribution is similar to the Fermi-Dirac distribution, and why the
hydrodynamic equations of \cite{csr}  involve a Fermi-Dirac-like pressure term.
However, the analogy
with quantum mechanics stops here because the Lynden-Bell theory is not
based on a Schr\"odinger equation. As a result, there is no quantum
potential (corresponding to the kinetic term, accounting for the Heisenberg
uncertainty principle, in the Schr\"odinger equation). There has been some
attempts to describe the Vlasov-Poisson equations on a coarse-grained scale in
terms of an effective Schr\"odinger equation \cite{wk} but this is
essentially an heuristic procedure aimed at smoothing out the small scales and
avoiding
numerical instabilities (see the discussion in \cite{prd3}).}
On
the other hand,  we have argued in this paper that a collisionless
self-gravitating system
evolving in a nondifferentiable spacetime (it could also be a classical system
having a chaotic motion on a very long timescale) is described by the
generalized Schr\"odinger equation (\ref{we6}). The corresponding hydrodynamic
equations (\ref{tmad4})-(\ref{tmad9}) have the form of damped
Euler equations including a quantumlike
potential (arising from the nondifferentiability of spacetime) and a temperature
term. However, there is no Lynden-Bell (Fermi-Dirac) pressure term in these
equations. In order to
reconcile the two descriptions (i.e. to recover, in the differentiable limit,
the hydrodynamic equations of \cite{csr} associated with the Lynden-Bell theory)
we derive in Appendix \ref{sec_lb} a generalized Schr\"odinger equation with a
nonlinearity accounting for Lynden-Bell's degeneracy
pressure. This may be seen as a
refinement of the generalized Schr\"odinger equation (\ref{we6}) taking into
account the Lynden-Bell exclusion principle of violent relaxation or,
alternatively, a generalization of the Lynden-Bell theory in a
nondifferentiable spacetime.

Let us briefly summarize the main ideas of this paper. We have proposed
that DM halos are described by the generalized Schr\"odinger equation
(\ref{we6}). This Schr\"odinger-like equation is not due to ordinary
quantum mechanics requiring ultralight bosons (like in the standard BECDM model)
but to the fractal (nondifferentiable) nature of spacetime that manifests
itself at astrophysical and cosmological scales. If true, this means that we
must take into account, in any theory of astrophysics and cosmology, the fractal
(nondifferentiable) nature of spacetime at large
scales, leading to a new form of quantum mechanics. We have suggested that
this new quantum mechanics
solves the problems of the CDM model just like the
standard BECDM model. However, it allows for a wider class of DM particles,
not only ultralight bosons. We have furthermore introduced the more general
Schr\"odinger equations (\ref{sr4}) and (\ref{lb5}) taking into account the
self-interaction of the particles or the  process of
violent
relaxation (Lynden-Bell's exclusion principle) in a fractal (nondifferentiable)
spacetime.

\appendix

\section{Short-range interactions}
\label{sec_sr}

In this Appendix, we generalize the Schr\"odinger
equation (\ref{we6}) by taking the self-interaction of the particles into
account.

\subsection{Mean-field Schr\"odinger equation}

Let us assume that the particles have a short-range interaction described by the
binary potential $u_{\rm SR}({\bf r}-{\bf r}')$. If we make a mean field
approximation, we find that the potential in which a particle moves is given by 
\begin{equation}
\label{sr1}
\Phi_{\rm SR}({\bf r},t)=\int u_{\rm SR}({\bf r}-{\bf r}')\rho({\bf r}',t)\,
d{\bf r}'.
\end{equation}
This short-range potential has to be added in Eq. (\ref{we6})
to the gravitational potential
corresponding to long-range interactions. The generalized  Schr\"odinger
equation
taking into account both short-range and long-range interactions in the mean
field approximation is
\begin{eqnarray}
\label{sr2}
i{\cal D}\frac{\partial\psi}{\partial
t}=-{\cal D}^2\Delta\psi+\frac{1}{2}(\Phi+\Phi_{\rm SR})\psi+\frac{k_B
T}{m}\ln|\psi|\, \psi\nonumber\\
-\frac{1}{2}i\xi{\cal D}\left\lbrack \ln\left
(\frac{\psi}{\psi^*}\right )-\left\langle \ln\left (\frac{\psi}{\psi^*}\right
)\right\rangle\right\rbrack\, \psi,
\end{eqnarray}
where
\begin{eqnarray}
\Phi({\bf r},t)=-G\int \frac{\rho({\bf r}',t)}{|{\bf r}-{\bf r}'|}\,
d{\bf r}'
\end{eqnarray}
is the gravitational potential determined by the  Poisson equation
(\ref{poisson}) and
$\Phi_{\rm SR}({\bf r},t)$ is the short-range potential  given by Eq.
(\ref{sr1}).

\subsection{Gross-Pitaevskii-like equation}

If we consider a pair contact interaction $u_{\rm
SR}=g\delta({\bf r}-{\bf
r}')$ with strength $g$ as described by
Dirac's $\delta$-function, we find
that 
\begin{equation}
\label{sr3}
{\Phi}_{\rm SR}({\bf r},t)=g \rho({\bf r},t).
\end{equation}
Using Eq. (\ref{jeu}), the foregoing equation can be rewritten as
$\Phi_{\rm SR}=g|\psi|^2$. Substituting this relation into Eq. (\ref{sr2}), we
obtain a generalized 
Schr\"odinger equation of the form
\begin{eqnarray}
\label{sr4}
i{\cal D}\frac{\partial\psi}{\partial
t}=-{\cal
D}^2\Delta\psi+\frac{1}{2}\Phi\psi+\frac{1}{2}g|\psi|^2\psi+\frac{k_B
T}{m}\ln|\psi|\, \psi\nonumber\\
-\frac{1}{2}i\xi{\cal D}\left\lbrack \ln\left
(\frac{\psi}{\psi^*}\right )-\left\langle \ln\left (\frac{\psi}{\psi^*}\right
)\right\rangle\right\rbrack\, \psi.\qquad
\end{eqnarray}
It includes a cubic nonlinearity like in the Gross-Pitaevskii
equation \cite{gross1,gross2,gross3,pitaevskii1,pitaevskii2}. This is a
particular case of the generalized GP equation studied in
\cite{total}.
The corresponding hydrodynamic equations have the form of Eqs.
(\ref{tmad4})-(\ref{tmad6}) with an equation of state
\begin{eqnarray}
\label{sr5}
P=\rho\frac{k_B T}{m}+\frac{1}{2}g\rho^2.
\end{eqnarray}
Their equilibrium state describes DM halos with a polytropic core
(soliton) of index $\gamma=2$ and an isothermal atmosphere. The polytropic
equation of state introduces an internal energy
$U=(1/2)g\int\rho^2\, d{\bf r}$
in the expression of the free energy (see Sec. \ref{sec_h} and
Ref. \cite{total}). We note that $U=(1/2)\int\rho\Phi_{\rm
SR}\, d{\bf r}$ where $\Phi_{\rm
SR}$ given by Eq. (\ref{sr3}).

{\it Remark:} in the quantum model where the bosons have a self-interaction, Eq.
(\ref{sr4}) takes the form
\begin{eqnarray}
\label{sr4qw}
i\hbar\frac{\partial\psi}{\partial
t}=-\frac{\hbar^2}{2m}\Delta\psi+m\Phi\psi+\frac{4\pi a_s\hbar^2}{m^2}
|\psi|^2\psi\nonumber\\
+2 k_B
T\ln|\psi|\, \psi
-i\frac{\hbar}{2}\xi\left\lbrack \ln\left
(\frac{\psi}{\psi^*}\right )-\left\langle \ln\left (\frac{\psi}{\psi^*}\right
)\right\rangle\right\rbrack\, \psi,\nonumber\\
\end{eqnarray}
where $a_s$ is the s-scattering length of the bosons and we
have used $g=4\pi a_s\hbar^2/m^3$ \cite{revuebec}. This corresponds to the
model of BECDM proposed in \cite{total}. The transition between
the region dominated by the quantum potential and the region dominated by
(effective) thermal effects corresponds to the (effective) de
Broglie length
\begin{eqnarray}
\lambda_{\rm dB}\sim
\frac{\hbar}{\sqrt{mk_B T}}.
\end{eqnarray}
The quantum potential
dominates for $r\ll \lambda_{\rm dB}$ and the effective thermal effects
dominate for   $r\gg \lambda_{\rm dB}$. The transition between
the region dominated by the quantum pressure (due to the repulsive
self-interaction of the bosons) and the region dominated by the
(effective) thermal pressure corresponds to a density
\begin{eqnarray}
\rho_B \sim \frac{m^2 k_B T}{a_s\hbar^2}.
\end{eqnarray}
The quantum pressure dominates for $\rho\gg\rho_B$ and the (effective) thermal
pressure dominates for  $\rho\ll\rho_B$.

\subsection{Cahn-Hilliard-like equation}

The potential of Eq. (\ref{sr3}) corresponds to the
dominant term in an expansion of the short-range potential of interaction
(\ref{sr1}) in powers of the range of the interaction. Let us derive a
generalized Schr\"odinger equation taking into account the next order term
in this expansion. Setting ${\bf
q}={\bf r}'-{\bf r}$ and writing the short-range potential of
interaction as
\begin{equation}
\label{ch1}
{\Phi}_{\rm SR}({\bf r},t)=\int u_{\rm SR}(q){\rho}({\bf r}+{\bf q},t)\, d{\bf
q},
\end{equation}
we can Taylor expand ${\rho}({\bf r}+{\bf q},t)$ to second order in ${\bf q}$
to obtain
\begin{equation}
\label{ch2}
{\rho}({\bf r}+{\bf q},t)={\rho}({\bf
r},t)+\sum_{i}\frac{\partial{\rho}}{\partial
x_{i}}q_{i}+\frac{1}{2}\sum_{i,j}\frac{\partial^{2}{\rho}}{\partial
x_{i}\partial x_{j}}q_{i}q_{j}.
\end{equation}
Substituting this expansion into Eq. (\ref{ch1}), we get
\begin{equation}
\label{ch3}
{\Phi}_{\rm SR}({\bf r},t)=g \rho({\bf r},t)+\chi\Delta\rho({\bf r},t)
\end{equation}
with $g=4\pi\int_{0}^{+\infty} u_{\rm SR}(q) q^{2} dq$ and $\chi=\frac{2\pi}{3}
\int_{0}^{+\infty} u_{\rm SR}(q) q^{4} dq$. Note that $l=(\chi/g)^{1/2}$
has the dimension of a length corresponding to the range of the
interaction.  Using Eq.
(\ref{jeu}), Eq. (\ref{ch3}) can be rewritten as ${\Phi}_{\rm SR}=g
|\psi|^2+\chi \Delta |\psi|^2$.  Substituting this relation into
Eq. (\ref{sr2}), we obtain a generalized 
Schr\"odinger equation of the form
\begin{eqnarray}
\label{ch4}
i{\cal D}\frac{\partial\psi}{\partial
t}=-{\cal
D}^2\Delta\psi+\frac{1}{2}\Phi\psi
+\frac{1}{2}g|\psi|^2\psi+\frac{1}{2}\chi\Delta|\psi|^2 \psi\nonumber\\
+\frac{k_B
T}{m}\ln|\psi|\, \psi
-\frac{1}{2}i\xi{\cal D}\left\lbrack \ln\left
(\frac{\psi}{\psi^*}\right )-\left\langle \ln\left (\frac{\psi}{\psi^*}\right
)\right\rangle\right\rbrack\, \psi.\nonumber\\
\end{eqnarray}
It includes a cubic nonlinearity like in the Gross-Pitaevskii equation
\cite{gross1,gross2,gross3,pitaevskii1,pitaevskii2} and a
Laplacian term
like in the Cahn-Hilliard equation \cite{ch,bray} (this analogy will be
developed in a separate
paper). The cubic term introduces an internal energy $U=(1/2)g\int\rho^2\, d{\bf
r}$, and the Laplacian term introduces a square gradient energy
$W_{\chi}=-(1/2)\chi\int (\nabla\rho)^2\,
d{\bf r}$, in the expression of the free energy (see Sec.
\ref{sec_h} and
Ref. \cite{total}). We note that $U=(1/2)\int\rho\Phi_{\rm
SR}\, d{\bf r}$ where $\Phi_{\rm
SR}$ given by Eq. (\ref{ch3}).

\section{Violent relaxation in a nondifferentiable spacetime}
\label{sec_lb}

In a differentiable spacetime, a collisionless self-gravitating system is
described by the Vlasov-Poisson equations. The Vlasov-Poisson equations
experience a process of violent relaxation \cite{lb}. They are
expected to relax, on a coarse-grained scale, towards the Lynden-Bell
distribution:\footnote{In practice, the quasistationary state reached by the
system as a result of violent relaxation deviates
from the Lynden-Bell distribution because of
incomplete relaxation \cite{lb,csr,cg,dubrovnik}.} 
\begin{eqnarray}
\label{lb0}
\overline{f}({\bf r},{\bf
v})=\frac{\eta_0}{1+e^{\eta_0(v^2/2+\Phi({\bf
r})-\mu)/T_{\rm eff}}},
\end{eqnarray}
where $\eta_0$ is the maximum value of the fine-grained distribution function
and $T_{\rm eff}$ is an effective temperature (note that $T_{\rm
eff}$ has not the
dimension of a
temperature but $T_{\rm eff}/\eta_0$ has the dimension of a velocity
dispersion). The Lynden-Bell distribution is similar to
the
Fermi-Dirac distribution. In the nondegenerate limit
$\overline{f}\ll
\eta_0$, it takes a form similar to the Maxwell-Boltzmann
distribution. The evolution of  the coarse-grained
distribution function $\overline{f}({\bf
r},{\bf v},t)$ can be described by a
generalized Landau or Fokker-Planck equation \cite{csr,cg,dubrovnik} taking into
account
the Lynden-Bell exclusion principle $\overline{f}({\bf
r},{\bf v},t)\le \eta_0$ which is similar to the Pauli exclusion
principle in quantum mechanics (but with another
interpretation).\footnote{When coupled to the Poisson
equation, the Lynden-Bell distribution function yields a cluster with an
infinite mass because it does not take into account the
escape of high energy particles. An improved model with a finite mass, which
can be derived from the generalized Landau equation, is
provided by the fermionic King model \cite{cg,clm1,clm2}.}
From this kinetic equation, one can derive generalized
hydrodynamic equations
\cite{csr} that incorporate a pressure force with a Lynden-Bell  equation of
state (similar to the Fermi-Dirac equation of state in quantum mechanics)
and a linear friction
force. In the nondegenerate limit, the
Lynden-Bell equation of state reduces to
the isothermal equation of state, but with temperature
proportional to mass \cite{lb}. Indeed, the mass $m$ of the particles should not
occur in a
collisionless theory based on the Vlasov equation. Therefore, in
Lynden-Bell's theory, the isothermal equation of state writes
\begin{eqnarray}
\label{lb0iso}
P=\rho\frac{T_{\rm eff}}{\eta_0}.
\end{eqnarray}
The hydrodynamic equations derived in \cite{csr}
are similar to Eqs. (\ref{tmad4})-(\ref{tmad9}) except for the presence of the
quantum potential. In our approach, this term arises from the
nondifferentiability of spacetime. The generalized Schr\"odinger
equation (\ref{we6}), which is equivalent to Eqs. (\ref{tmad4})-(\ref{tmad9}),
may therefore describe the process of violent relaxation in a nondifferentiable
spacetime (or on a very long timescale when chaotic effects come into
play). To improve this description, one needs to take into account
the Lynden-Bell exclusion principle that is specific to the theory of violent
relaxation. Since the hydrodynamic equations (\ref{tmad4})-(\ref{tmad9}) already
contains a thermal pressure (it can be adapted to the theory of violent
relaxation by replacing $k_B T/m$ by $T_{\rm eff}/\eta_0$), we just have to add
the contribution of the
degeneracy pressure as explained below.

In the completely degenerate limit (corresponding to a system at $T_{\rm
eff}=0$), the
 Lynden-Bell distribution function is given by 
\begin{eqnarray}
\label{lb1}
\overline{f}({\bf r},{\bf v})=\eta_0 H(v_{\rm LB}({\bf r})-v), 
\end{eqnarray}
where $H(\cdot)$ is the Heaviside step function and $v_{\rm LB}({\bf r})$ is the
Lynden-Bell velocity (similar to the Fermi velocity in quantum mechanics).
The density and the pressure are
then given by
\begin{eqnarray}
\label{lb2}
\rho=\int \overline{f}\, d{\bf v}=\int_0^{v_{\rm LB}}\eta_0 4\pi v^2\,
dv=\frac{4\pi}{3}\eta_0 v_{\rm LB}^3,
\end{eqnarray}
\begin{equation}
\label{lb3}
P=\frac{1}{3}\int \overline{f}v^2\, d{\bf v}=\frac{1}{3}\int_0^{v_{\rm
LB}}\eta_0
v^2 4\pi v^2\, dv=\frac{4\pi}{15}\eta_0 v_{\rm LB}^5,
\end{equation}
leading to the equation of state\footnote{The general
Lynden-Bell equation of state $P_{\rm LB}(\rho)$, similar to the
Fermi-Dirac equation of
state in quantum mechanics, can be obtained by substituting Eq. (\ref{lb0}) into
the first integrals of Eqs. (\ref{lb2}) and (\ref{lb3}).}
\begin{eqnarray}
\label{lb4}
P=\frac{1}{5}\left (\frac{3}{4\pi\eta_0}\right )^{2/3}\rho^{5/3}.
\end{eqnarray}
This is a polytropic equation of state  of index $\gamma=5/3$ ($n=3/2$) like in
the theory of white dwarf stars \cite{chandra}. 

It is easy to determine the new term in the generalized
Schr\"odinger equation that leads to an
equation of state  of that form \cite{total}. We find 
\begin{eqnarray}
\label{lb5}
i{\cal D}\frac{\partial\psi}{\partial
t}=-{\cal D}^2\Delta\psi+\frac{1}{2}\Phi\psi+\frac{1}{4}\left
(\frac{3}{4\pi\eta_0}\right )^{2/3}|\psi|^{4/3}\psi\nonumber\\
+\frac{T_{\rm eff}}{\eta_0}\ln|\psi|\, \psi
-\frac{1}{2}i\xi{\cal D}\left\lbrack \ln\left
(\frac{\psi}{\psi^*}\right )-\left\langle \ln\left (\frac{\psi}{\psi^*}\right
)\right\rangle\right\rbrack\, \psi.\nonumber\\
\end{eqnarray}
This generalized Schr\"odinger equation includes a power-law nonlinearity 
$|\psi|^{4/3}\psi$ that
generalizes the one arising in the Gross-Pitaevskii equation. The
corresponding
hydrodynamic equations have the form of Eqs. (\ref{tmad4})-(\ref{tmad6}) with an
equation of state
\begin{eqnarray}
\label{lb6}
P=\rho\frac{T_{\rm eff}}{\eta_0}+\frac{1}{5}\left (\frac{3}{4\pi\eta_0}\right
)^{2/3}\rho^{5/3}.
\end{eqnarray}
Their equilibrium state describes DM halos with a solitonic
core due to the quantumlike potential (arising from the nondifferentiability
of spacetime), a polytropic
core
(similar to a fermion ball) of index $n=3/2$ due to
Lynden-Bell's type of degeneracy, and an isothermal
atmosphere due to violent relaxation or being a manifestation of
the temperature of an aether (it is also possible to take into account the
self-interaction of the particles as in Appendix \ref{sec_sr}).
The polytropic
equation of state introduces an internal energy $U=(3/10)\left
({3}/{4\pi\eta_0}\right )^{2/3}\int \rho^{5/3}\, d{\bf r}$ 
in the expression of the free energy (see Sec. \ref{sec_h}
and Ref. \cite{total}). This expression can
be directly obtained from the relation $U=\frac{1}{2}\int \overline{f}
v^2\, d{\bf r}d{\bf v}=\frac{3}{2}\int P\, d{\bf r}$, where $P$ is the
``quantum'' pressure at $T_{\rm eff}=0$ given by Eq.
(\ref{lb4}). In other words, the internal energy $U$
corresponds to the kinetic energy of ``microscopic'' motions.

The hydrodynamic equations (\ref{tmad4})-(\ref{tmad6})
corresponding to the generalized Schr\"odinger equation (\ref{lb5}) are similar
to the hydrodynamic equations obtained in \cite{csr} except that they include
a quantum potential arising from the nondifferentiability
of spacetime.\footnote{In the present approach, the Lynden-Bell
equation of state $P_{\rm LB}(\rho)$ appearing in the hydrodynamic equations of
\cite{csr}  is replaced by the simpler equation of state (\ref{lb6}). It is
easy to determine the nonlinear term $h(|\psi|^2)$ in the generalized
Schr\"odinger equation that exactly reproduces the Lynden-Bell equation of
state $P_{\rm LB}(\rho)$. It
corresponds to the associated enthalpy $h_{\rm LB}(\rho)=\int^{\rho}
[P'_{\rm LB}(\rho')/\rho']\, d\rho'$ \cite{total}. However, since it does not
have an
analytical
expression, we only consider here the simpler equation of state (\ref{lb6}).}
As a result, they can be viewed as a generalization of the
equations of violent relaxation \cite{csr} in a fractal (nondifferentiable)
spacetime. Alternatively, Eq. (\ref{lb5}) can be viewed as a refinement of the
generalized
Schr\"odinger equation
(\ref{we6}) taking into account the specificities of violent relaxation
(Lynden-Bell's exclusion principle).

In summary, the hydrodynamic equations 
(\ref{tmad4})-(\ref{tmad6}) with the equation of state (\ref{lb6}) and with the
quantum potential describe the process of violent relaxation in a
nondifferentiable spacetime while the hydrodynamic equations
(\ref{tmad4})-(\ref{tmad6}) with the equation of state  (\ref{lb6}) but without
the quantum potential describe the process of violent relaxation in a
differentiable spacetime.

{\it Remark:} In the case of BECDM where the particles are
bosons and where the nondifferentiability of spacetime is due
to quantum mechanics, using Eq. (\ref{qm3}), we find that Eq. (\ref{lb5})
becomes 
\begin{eqnarray}
\label{lb5Q}
i\hbar\frac{\partial\psi}{\partial
t}=-\frac{\hbar^2}{2m}\Delta\psi+m\Phi\psi+\frac{m}{2}\left
(\frac{3}{4\pi\eta_0}\right )^{2/3}|\psi|^{4/3}
\psi\nonumber\\
+\frac{2m}{\eta_0} T_{\rm eff} \ln|\psi|\, \psi
-i\frac{\hbar}{2}\xi \left\lbrack \ln\left
(\frac{\psi}{\psi^*}\right )-\left\langle \ln\left (\frac{\psi}{\psi^*}\right
)\right\rangle\right\rbrack\, \psi.\nonumber\\
\end{eqnarray}
This equation generalizes the BECDM model by taking into
account the process of violent relaxation.
This description may also
be valid if the  particles are fermions at statistical
equilibrium. In that
case, the isothermal equation of state corresponds to the true thermodynamic
temperature, the polytropic equation of state takes into account the Pauli
exclusion principle, and the quantum potential takes into
account the Heisenberg uncertainty principle. This leads to a generalized 
Schr\"odinger equation of the form \cite{total}:\footnote{It
may also be relevant to include the Dirac-Slater \cite{diracHF,slater} exchange
correction arising from the identity
of the fermions as in Sec. 5.2 of \cite{total}.}
\begin{eqnarray}
\label{lb5si}
i\hbar\frac{\partial\psi}{\partial
t}=-\frac{\hbar^2}{2m}\Delta\psi+m\Phi\psi+\frac{1}{2}\left
(\frac{3}{8\pi}\right )^{2/3}\frac{(2\pi
\hbar)^2}{m^{5/3}}|\psi|^{4/3}\psi\nonumber\\
+2k_B T \ln|\psi|\, \psi
-i\frac{\hbar}{2}\xi \left\lbrack \ln\left
(\frac{\psi}{\psi^*}\right )-\left\langle \ln\left (\frac{\psi}{\psi^*}\right
)\right\rangle\right\rbrack\, \psi.\nonumber\\
\end{eqnarray}
The other relevant equations can be obtained from the previous ones by using
the relation $\eta_0=2m^4/(2\pi\hbar)^3$ \cite{chandra}. The
transition between
the region dominated by the quantum pressure (due to the Pauli exclusion
principle) and the region dominated by the thermal pressure corresponds to a
density
\begin{eqnarray}
\rho_F \sim \frac{m^{5/2}(k_B T)^{3/2}}{\hbar^3}.
\end{eqnarray}
The quantum pressure dominates for $\rho\gg\rho_F$ and the thermal
pressure dominates for  $\rho\ll\rho_F$.

\section{Effective temperature}
\label{sec_tef}

We have seen that large DM halos have an isothermal, or almost isothermal, 
atmosphere which is responsible for the flat, or almost flat, rotation
curves of galaxies. The temperature $T$ is related to the circular velocity at
infinity $v_{\infty}$ by the relation
\begin{equation}
\frac{k_B T}{m}=\frac{v_{\infty}^2}{2}.
\label{effT}
\end{equation}
For the Medium Spiral,
$v_{\infty}\sim 153\, {\rm km/s}$. If we assume that $T$ is the
true thermodynamic temperature, then $m$ represents the mass of the DM
particle and  Eq. (\ref{effT}) determines the temperature of the halo $T$ as a
function of the mass of the DM particle.

If we assume that DM halos are self-gravitating BECs, then the boson mass must
be of the order of  $m=2.92\times 10^{-22}\, {\rm eV}/c^2$ in order to account
for the mass and size of ultracompact dwarf halos at $T=0$ such as Fornax (see
Sec. \ref{sec_coeffD} and Appendix D of \cite{abrilphas}). In that case, we find
from Eq. (\ref{effT}) that the temperature of large halos such as the Medium
Spiral
is $T\sim 4.41\times 10^{-25}\, {\rm K}$.\footnote{Bosons with a
repulsive
self-interaction may have a much larger mass than noninteracting bosons, up to
$m=1.10\times 10^{-3}\, {\rm
eV}/c^2$ (see
Appendix D of \cite{abrilphas}),  leading to a temperature $T\sim 1.66\times
10^{-6}\, {\rm K}$ (we have $T\sim 1\, {\rm K}$ for
$m\sim 662\, {\rm eV/c^2}$).} Such a small temperature may not be
physical.\footnote{Actually, we can
take the opposite point of view. We can argue that the temperature of the aether
$T\sim 4.41\times 10^{-25}\, {\rm K}$ is physical but it is so small that it is
undetectable in earth experiments. However, if the mass of the particle
is extraordinarily small, such as $m=2.92\times 10^{-22}\, {\rm eV}/c^2$, the
ratio $k_B T/m$ becomes large and can have observable consequences such as on
the rotation curves of the galaxies in astrophysics. Similarly, the friction
coefficient $\xi$ with
the aether is undetectable in earth
experiments (the friction time $\xi^{-1}\sim 1\, {\rm Myrs}$ is extremely long)
but it becomes important on astrophysical and cosmological
timescales. We can also make the following remark. In the present point of
view, the temperature of the aether is $T\sim 4.41\times 10^{-25}\, {\rm K}$ at
the scale
$\sim 10\, {\rm kpc}$ corresponding to the Medium Spiral while $T=0$ at the
scale $\sim 1\, {\rm kpc}$ corresponding to the ground state  (Fornax) of the
BECDM model (see Sec. \ref{sec_coeffD}). At smaller scales, $T$ could
become negative in order to
recover the results of Bialynicki-Birula \& Mycielski \cite{bm} (gaussons)  at
the microscale. This is discussed in more detail in Sec. 7 of \cite{total}.}
This strongly
suggests that $T$ is not the true thermodynamic temperature.\footnote{The
bosons would have an isothermal equation of state $P=\rho k_B T/m$ (due to
thermodynamics) if $T\gg T_c$, where $T_c$ is their condensation temperature
given by $T_c=2\pi \hbar^2\rho^{2/3}/[m^{5/3}k_B\zeta(3/2)^{2/3}]$ with
$\zeta(3/2)=2.6124...$ Evaluated for large halos such as the Medium
Spiral where $\rho_0=7.02\times 10^{-3}M_{\odot}/{\rm
pc}^3$,  we get $T_c=4.82\times 10^{36}\, {\rm K}$ (!) It is considerably larger
than $T\sim 4.41\times 10^{-25}\, {\rm K}$ (and than any
reasonable temperature) indicating that the bosons are
completely condensed and that we can consider that $T_{\rm thermo}=0$.}  It may
rather
represent an effective temperature. We have
proposed two possible
interpretations of this effective temperature:

(i) We have suggested that the quantum-like aspects of DM halos
are not due to quantum mechanics but to the fractal structure of spacetime at
the cosmic scale. In that case,  DM halos are described by a generalized
Schr\"odinger-like equation
\begin{eqnarray}
\label{we4bis}
i{\cal D}\frac{\partial\psi}{\partial
t}=-{\cal
D}^2\Delta\psi+\frac{1}{2}\Phi\psi+\frac{1}{2}V(t)\psi\nonumber\\
+\frac{v_{\infty}^2}{2}\ln|\psi|\, \psi-\frac{1}{2}i\xi{\cal D}\ln\left
(\frac{\psi}{\psi^*}\right )\, \psi,
\end{eqnarray}
where neither the mass of the DM particle nor the temperature appear
explicitly. The flat rotation curves of galaxies is due to the term $
(v_{\infty}^2/2)\ln|\psi|\, \psi$ in the generalized
Schr\"odinger-like equation (\ref{we4bis}), where $v_{\infty}$ is a coefficient
of this
equation. It could be interpreted as a sort of fundamental constant of physics,
except that it depends on the scale as discussed in Sec. \ref{sec_coeffT}. In
this interpretation, there is no ultra-small mass $m$ nor ultra-small
temperature $T$ since such quantities do not  explicitly appear in the
equations.
One can always define a temperature $T$ by the relation $v_{\infty}^2/2=k_B
T/m$ (where $m$ is some mass scale) in order to develop a thermodynamical
analogy, but this temperature is purely effective since only the ratio $k_B
T/m$ has a physical meaning. It could be interpreted as the
temperature of the aether but the process of thermalization would be completely
different than in thermodynamics.

(ii)  We have suggested that the envelope of DM halos arises from a process of
collisionless violent relaxation like in the theory of Lynden-Bell \cite{lb}. 
Such a process tends to establish an isothermal distribution justifying Eq.
(\ref{effT}). However, since this process is based on the Vlasov equation, the
mass of the DM particle should not appear in the equations. In other words, 
the temperature $T$ must be proportional to mass \cite{lb}. In
Lynden-Bell's theory, $k_B T/m$ is replaced by $T_{\rm eff}/\eta_0$ where
$T_{\rm eff}$ is an effective temperature (see Appendix \ref{sec_lb}). 
Therefore $v_{\infty}^2/2=T_{\rm eff}/\eta_0$. Again, in
this
interpretation, there is no ultra-small mass $m$ nor ultra-small
temperature $T$ since such quantities do not  explicitly  appear in the
equations.

{\it Remark:} If we assume that DM halos are made of
fermions, like a sterile neutrino, then the fermion mass must be of the
order of  $m=170\, {\rm eV}/c^2$ in order to account
for the mass and size of ultracompact dwarf halos at $T=0$ such as Fornax (see
Appendix D of \cite{abrilphas}). In that case, we find from Eq. (\ref{effT})
that the temperature of large halos such as the Medium Spiral
is $T\sim 0.257\, {\rm K}$. This temperature is physical (and there is no
condensation temperature in the Fermi-Dirac statistics) suggesting that,
if DM is made of fermions, $T$ may represent the true thermodynamic
temperature.

\section{Generalized Einstein relation}
\label{sec_einstein}

In this Appendix, we derive the time-independent generalized
Schr\"odinger equation (\ref{app1}) from the method of quantization introduced
by Schr\"odinger in his first paper on quantum mechanics \cite{schrodinger1}. In
this paper, he derived the fundamental eigenvalue equation of quantum mechanics
from a variational principle based on  the classical Hamilton-Jacobi equation
(see Appendix F of \cite{chavmatos} for a short account of his approach). We use
the same approach but take into account frictional effects. 

The classical
Hamilton-Jacobi equation with friction is
\begin{equation}
\label{einstein1}
\frac{\partial \sigma}{\partial t}+\frac{(\nabla
\sigma)^2}{2}+\Phi+\xi\sigma=0,
\end{equation}
where $\sigma$ is the classical action which is related to the classical
velocity by ${\bf
u}=\nabla \sigma$. Introducing the energy
$\epsilon=-\partial\sigma/\partial t$, we get 
\begin{equation}
\label{einstein2}
\epsilon=\frac{(\nabla
\sigma)^2}{2}+\Phi+\xi\sigma.
\end{equation}
Following Schr\"odinger's approach, we introduce a real wave function $\phi({\bf
r})$ through the substitution
\begin{equation}
\label{einstein3}
\sigma=2{\cal D}\ln\phi.
\end{equation}
Equation (\ref{einstein2}) is then rewritten in terms of $\phi$ as
\begin{equation}
\label{einstein4}
(\nabla\phi)^2-\frac{1}{2{\cal D}^2}(\epsilon-\Phi)\phi^2+\frac{\xi}{\cal
D}(\ln\phi)\phi^2=0.
\end{equation}
Following again Schr\"odinger's approach, we introduce the functional
\begin{equation}
\label{einstein5}
J=\int \left\lbrace (\nabla\phi)^2-\frac{1}{2{\cal
D}^2}(\epsilon-\Phi)\phi^2+\frac{\xi}{\cal
D}(\ln\phi)\phi^2
\right\rbrace\, d{\bf r}
\end{equation}
and consider its minimization with respect to variations on $\phi$. The
stationarity condition $\delta J=0$ gives
\begin{equation}
\label{einstein6}
-2{\cal D}^2\Delta\phi+\Phi\phi+2\xi{\cal D} (\ln\phi)\phi=\tilde\epsilon\phi,
\end{equation}
where we have redefined the energy as $\tilde\epsilon=\epsilon-\xi{\cal D}$ for
convenience. Comparing this equation with the time-independent generalized
Schr\"odinger equation (\ref{app1}), we obtain the relation
\begin{eqnarray}
\label{einstein7}
{\cal D}=\frac{k_B T}{m\xi}.
\end{eqnarray}
In the case of quantum mechanics, using Eq. (\ref{qm3}), it takes the form
\begin{eqnarray}
\label{einstein7b}
\frac{\hbar}{2m}=\frac{k_B T}{m\xi}\qquad {\rm or}\qquad
\frac{\hbar}{2}=\frac{k_B T}{\xi}.
\end{eqnarray}
This can be viewed as a sort of generalized Einstein relation expressing a form
of  fluctuation-dissipation theorem. This relation can also be obtained from the
formalism of scale relativity by assuming that $\gamma_R=\gamma_I$ (see Appendix
G of \cite{chavnot}). Therefore, in a sense, the variational approach of
Schr\"odinger applied to the present situation suggests that
\begin{eqnarray}
\label{einstein8}
\gamma_R=\gamma_I.
\end{eqnarray}
We do not claim, however, that this equality should always be valid.

\end{document}